\documentclass[aps,prb,twocolumn,superscriptaddress,showpacs]{revtex4-2}
\usepackage{graphicx,bm,color,textcomp,mathtools,times,nicefrac,soul,xcolor}
\usepackage{ulem}
\usepackage{enumitem}
\usepackage{appendix}
\newcommand\Nd{$\rm Nd_2Zr_2O_7$}
\newcommand\Cer{$\rm Ce_2Zr_2O_7$}
\newcommand\Pra{$\rm Pr_2Zr_2O_7$}

\newcommand\HIIO{$H \! \parallel \! [110]$}
\newcommand\HIOO{$H \! \parallel \! [100]$}

\setstcolor{magenta}

\begin{document}
	\title{Large out-of-equilibrium magnetocaloric effect
    in rare-earth zirconate pyrochlores}
	\date{\today}
	\author{O.~Benton}		
    \email{j.o.benton@qmul.ac.uk}
    \affiliation{School of Physical and Chemical Sciences,
    Queen Mary University of London, London E1 4NS, 
    United Kingdom}
    \affiliation{Max Planck Institute for the Physics of Complex Systems, N\"{o}thnitzer Str. 38, Dresden 01187, Germany}
	\author{Y.~Skourski}
    \affiliation{Hochfeld-Magnetlabor Dresden (HLD-EMFL) and W\"urzburg-Dresden Cluster of Excellence ct.qmat, Helmholtz-Zentrum Dresden-Rossendorf, 01328 Dresden, Germany}
	\author{D.~Gorbunov}			\affiliation{Hochfeld-Magnetlabor Dresden (HLD-EMFL) and W\"urzburg-Dresden Cluster of Excellence ct.qmat, Helmholtz-Zentrum Dresden-Rossendorf, 01328 Dresden, Germany}
	\author{A.~Miyata}				\affiliation{Hochfeld-Magnetlabor Dresden (HLD-EMFL) and W\"urzburg-Dresden Cluster of Excellence ct.qmat, Helmholtz-Zentrum Dresden-Rossendorf, 01328 Dresden, Germany}
	\author{S.~Chattopadhyay}		\affiliation{Hochfeld-Magnetlabor Dresden (HLD-EMFL) and W\"urzburg-Dresden Cluster of Excellence ct.qmat, Helmholtz-Zentrum Dresden-Rossendorf, 01328 Dresden, Germany}
	\author{J.~Wosnitza}			\affiliation{Hochfeld-Magnetlabor Dresden (HLD-EMFL) and W\"urzburg-Dresden Cluster of Excellence ct.qmat, Helmholtz-Zentrum Dresden-Rossendorf, 01328 Dresden, Germany}
                                    \affiliation{Institut f\"ur Festk\"orper- und Materialphysik, Technische Universit\"at Dresden, 01062 Dresden, Germany}
	\author{M.~Ciomaga~Hatnean}
% \altaffiliation[Current addresses: ]{PSI Center for Neutron and Muon Sciences, Paul Scherrer Institute, 5232 Villigen PSI, Switzerland \& Materials Discovery Laboratory, Department of Materials, ETH Z\"urich, 8093 Z\"urich, Switzerland}
								\affiliation{Department of Physics, University of Warwick, Coventry CV4 7AL, United Kingdom}
    \author{G.~Balakrishnan}		\affiliation{Department of Physics, University of Warwick, Coventry CV4 7AL, United Kingdom}
	\author{S.~Zherlitsyn}			\affiliation{Hochfeld-Magnetlabor Dresden (HLD-EMFL) and W\"urzburg-Dresden Cluster of Excellence ct.qmat, Helmholtz-Zentrum Dresden-Rossendorf, 01328 Dresden, Germany}
	\author{O.~A.~Petrenko}			\email{O.Petrenko@warwick.ac.uk}
								    \affiliation{Department of Physics, University of Warwick, Coventry CV4 7AL, United Kingdom}	
\begin{abstract}	
We explore the magnetic properties of \Nd\ and \Pra\ single crystals subjected to pulsed magnetic fields up to 60 T using magnetization and magnetocaloric-effect (MCE) measurements, with initial temperatures ranging from 2 to 31~$\rm{K}$.
The MCE data exhibit pronounced and unconventional hysteresis loops, in which the sample temperature increases during both the up-sweep and down-sweep of the field.
In \Nd, the MCE further displays a striking plateau as a function of time, followed by a rapid temperature rise that begins at the maximum applied field, across pulses with differing peak-field strengths.
Our magnetization measurements reveal an inferred temperature of the magnetic subsystem that differs significantly from the directly measured sample temperature and exhibits opposite hysteresis: the temperature is higher on the up-sweep than the down-sweep, unlike the direct measurements.
These observations indicate a breakdown of thermal equilibrium between magnetic and lattice degrees of freedom on the timescale of the pulse ($\sim 10^{-1}$~s).
We interpret the results using a phenomenological model involving two thermally coupled subsystems—the magnetic ions and phonons, and a thermal reservoir, which accounts well for the behavior of \Pra. However, it fails to reproduce the plateau seen in \Nd.
Agreement with \Nd\ data is improved substantially if we allow the thermal coupling between the magnetic and the lattice subsystems to depend on the product $\frac{HdH}{dt}$.
Our results reveal anomalously slow heat transfer between magnetic and lattice subsystems and point toward a novel mechanism for dynamically controlling the heat flow in \Nd\ via the rate of magnetic field variation.
\end{abstract}

\maketitle

\section{Introduction}
Over the past two decades, several members of the zirconate pyrochlore family, $Ln_2\mathrm{Zr_2O_7}$ (with $Ln=$ Ce, Nd, Pr, Sm, Gd), have become the focus of intense experimental investigation~\cite{Rau_2019, Gao_2019, Gaudet_2019, Xu_2015, Ciomaga_2015, Kimura_2013, Kimura_2013_JKPS, Bonville_2016,
Singh_2008, Xu_2017_Thesis}.
These studies aimed to uncover manifestations of unconventional magnetic behavior arising from the distinct quantum nature of the rare-earth ions and their interactions.
Major developments include the observation of magnetic fragmentation in \Nd~\cite{Petit_2016, Benton_2016}, signatures of dynamic kagome ice~\cite{Lhotel_2018}, disorder-induced quantum paramagnetism in \Pra~\cite{Wen_2017, Benton_2018}, and a liquid-gas-like metamagnetic transition in the same compound~\cite{Tang_2023}, as well as a proposed $U(1)_\pi$ quantum spin-liquid ground state in the dipole-octupole compound \Cer~\cite{Smith_2022, Gao_2022, Gao_arxiv, Smith_2024}.
These findings highlight the rich and unconventional magnetism that emerges in these geometrically frustrated systems.

The diversity of phenomena observed at zero or low magnetic fields naturally raises the question of how these materials respond to strong magnetic fields.
In this article, we address this by reporting measurements of the magnetic and thermodynamic properties of \Nd\ and \Pra\ in pulsed magnetic fields up to 60~T.
Specifically, we present magnetization $M(H)$ and magnetocaloric effect (MCE) $T(H)$ data obtained from a series of pulsed-field measurements, with peak-field strengths ranging from 5 to 60~T, with fields oriented along the $[100]$, $[110]$, and $[111]$ directions.

We observe hysteresis loops in the magnetization of both compounds -- an unexpected result, given that our measurements are performed at temperatures well above any putative magnetic-ordering temperature.
We also observe pronounced hysteresis in the magnetocaloric effect, with an unconventional character: the sample temperature continues to {\it increase} even during the {\it decreasing} portion of the field pulse, reaching its maximum when the applied field has fallen to as little as half or even a quarter of its peak value.

At first glance, the hysteresis observed in the $M(H)$ and $T(H)$ curves appears inconsistent.
Whereas the directly measured temperature is higher on the down-sweep of the field, the temperature inferred from magnetization measurements is higher on the up-sweep.
These observations collectively point to a strong breakdown of thermal equilibrium between the magnetic and lattice degrees of freedom (DOFs).

The field-dependent temperature measurements in \Nd\ reveal a further surprising feature when plotted as a function of time, $T(t)$.
For the strongest field pulses, we observe a pronounced plateau in temperature, followed by a rapid increase.
This temperature rise begins at the point where the applied field reaches its maximum value.
Strikingly, this behavior persists across pulses with different peak-field strengths.

The measurement of the MCE in pulsed magnetic fields is a well-established and widely used technique, employed routinely at several large-scale facilities~\cite{Kihara_2013,Brambleby_2017}.
However, we are not aware of any previous studies reporting such unusual $T(H)$ behavior in other materials, in the absence of field-induced phase transitions.

Various pyrochlore magnets have previously been studied in high and ultrahigh magnetic fields~\cite{Opherden_2019,Erfanifam_2014,Tang_2024}, where unexpected field-induced behavior has often been attributed to the interplay between exchange interactions and strong single-ion anisotropies.
In the case of \Nd\ and \Pra\ studied here, the situation is markedly different: within the temperature range of our measurements, exchange interactions play a negligible role, and the physics is dominated by the single-ion response.

To rationalize our observations, we introduce a phenomenological model consisting of two weakly coupled subsystems: magnetic DOFs and lattice DOFs, as well as a thermal reservoir.
With constant coupling strengths, the model provides a good description of the behavior observed in \Pra.

For \Nd, however, the same model fails to reproduce the most striking feature of the data—the temperature plateau followed by a rapid rise in $T(t)$.
To describe the \Nd\ data, we introduce an explicit dependence of the thermal coupling between magnetic and lattice DOFs on the rate of change of field intensity $H \frac{d H}{d t} = \frac{1}{2} \frac{d (H^2)}{d t}$.
This substantially improves the agreement with the experimental data.

The paper is organized as follows.
Section~\ref{sec:experimental} describes the experimental techniques used in this study.
Sections~\ref{sec:Pr} and ~\ref{sec:Nd}, each further divided into technique-specific subsections, present the results for \Pra\ and \Nd, respectively.
Section~\ref{sec:theory} introduces the phenomenological model, and Section~\ref{sec:summary} summarizes our conclusions.
A series of appendices provide additional details of the theoretical analysis.

%_________________________________________________________________________________________________
%_________________________________________________________________________________________________
\section{Experimental details} \label{sec:experimental}
Single-crystalline samples of \Nd\ and \Pra\ were grown by the floating-zone method using an infrared furnace as described in Ref.~\cite{Ciomaga_2016}. 
The crystals were examined and aligned using a backscattering X-ray Laue system.

High-field magnetization and MCE measurements were made using pulsed magnetic fields up to 60~T at the Dresden High Magnetic Field Laboratory (HLD), Germany.

The magnetization was measured using field pulses with a rise time of 7~ms and a total duration of 25~ms.
magnetization was obtained by integrating the voltage induced in a compensated coil system surrounding the sample~\cite{Skourski_2011}.
An empty magnetometer response was recorded and subtracted for each sample measurement.
With the samples shape being needle-like elongated along the direction of applied field, the demagnetizing fields could largely be neglected. 
The methods of converting the magnetization data into absolute units are described in the experimental results sections.

\begin{figure}[tb]
\includegraphics[width=0.9\columnwidth]{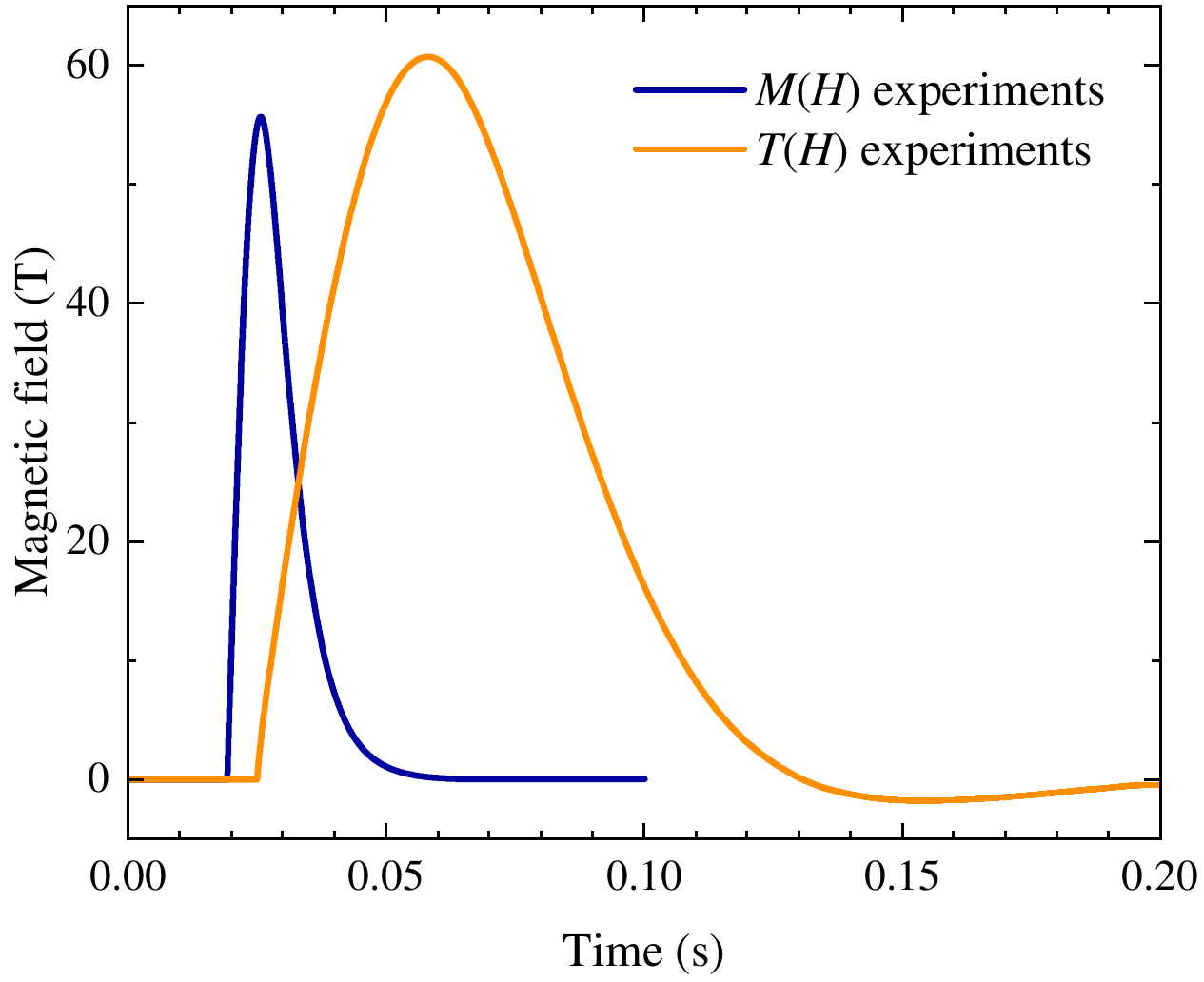}
\caption{Representative magnetic-field pulse profiles used in magnetization $M(H)$ and MCE $T(H)$ experiments.}
\label{fig:pulses}
\end{figure}

Direct temperature measurement, to obtain the MCE, was performed using miniature ($0.6 \times 0.3 \times 0.1$~mm$^3$) RuOx-based surface mounted device (SMD) resistors as thermometers.
At least two thermometers from the same batch were used -- one being attached to the sample and another coupled to the sample holder in order to record the magnetoresistance, which is taken into account during data evaluation.
The resistance of the SMD thermometers was measured using a numerical procedure within a lock-in technique at a frequency of 40~kHz and calibrated during the cool-down process against the system thermometer of the probe.

The magnetization and MCE experiments were conducted on different instruments with different field-pulse profiles.
Representatives of each pulse profile are shown in Fig.~\ref{fig:pulses}.
For the purpose of numerical calculations (see Sec.~\ref{sec:theory}), the field pulse is approximated by an interpolating function constructed in Python using the SciPy module {\it scipy.interpolate} \cite{scipy-ref}.

%_________________________________________________________________________________________________

\section{Experimental Results: \Pra} \label{sec:Pr}
\subsection{Magnetization measurements}	\label{sec:M_Pr}
\begin{figure}[b]
\includegraphics[width=0.95\columnwidth]{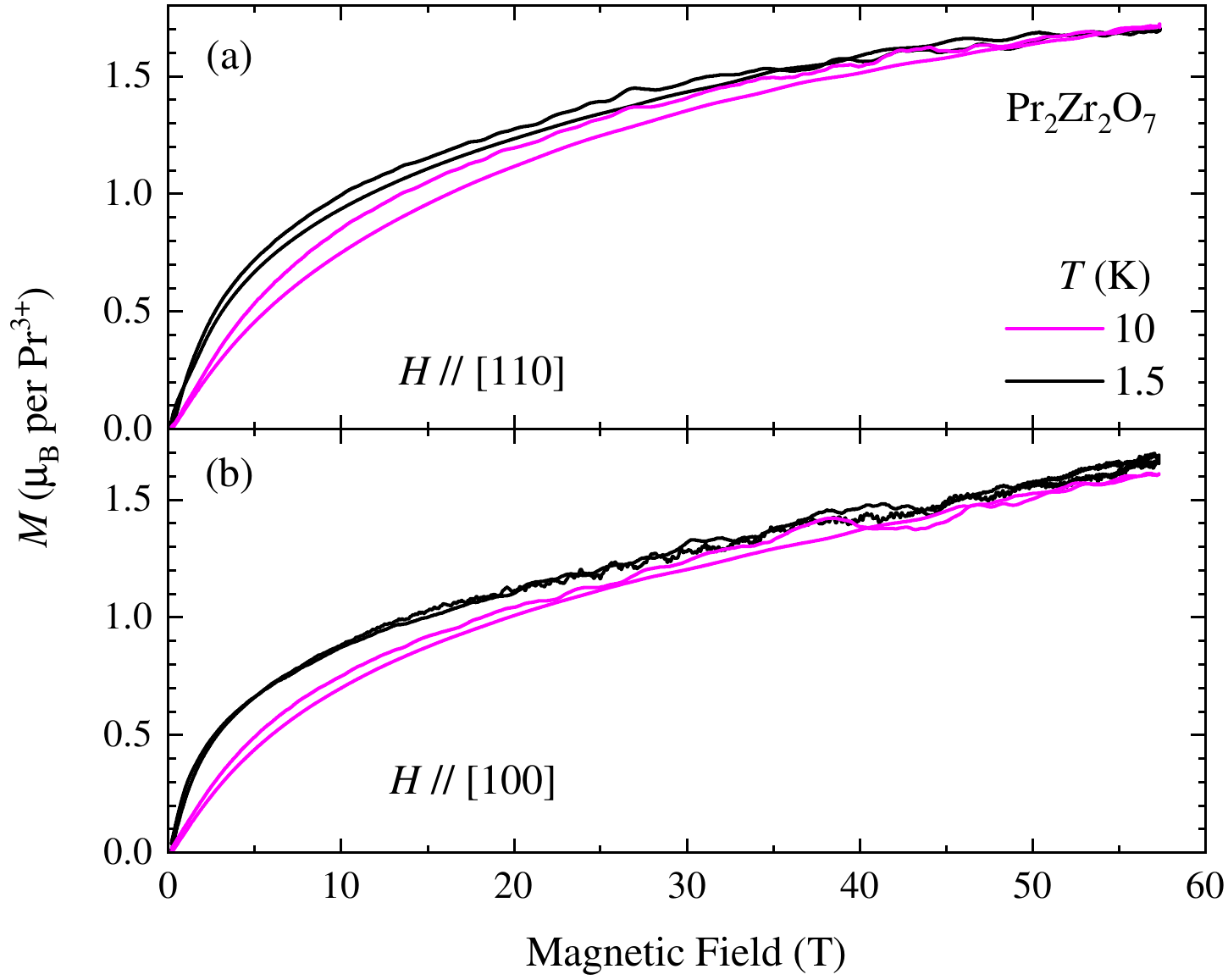}
\caption{Field dependence of the magnetization of \Pra\ for (a) \HIIO\ and (b) \HIOO\ at $T=1.5$ and 10~K.}
\label{fig:Pr_MH}
\end{figure}

The low-temperature magnetization process in \Pra\ has been reported in several publications~\cite{Kimura_2013_JKPS,Hatnean_2014,Petit_2016,Bonville_2016}, but the measurements were limited to relatively low fields (below 11~T).
Figure~\ref{fig:Pr_MH} presents high-field magnetization curves in \Pra\ for fields aligned parallel to [110] and [100].
The high-field magnetization data $M(H)$ were converted to absolute units by measuring low-field magnetization curves of the same samples using a conventional SQUID magnetometer.
One noticeable feature of the $M(H)$ curves is that, at high fields, they become almost isotropic, i.e., significant differences in the absolute value observed at low fields (in agreement with the previous measurements) gradually disappear with increasing field and at 57~T, the magnetization for \HIOO, an easy-axis direction, and \HIIO, a hard-axis direction, are practically identical.
It is also rather clear that the temperature dependence of magnetization is more pronounced in lower fields compared to high fields, as at 57~T, the magnetic moments are the same for the initial temperatures of 1.5 and 10~K.

Overall, the magnetization process in \Pra\ looks {\it conventional}, as the hysteresis observed is relatively small and there is a continuous increase and decrease of the magnetization for increasing and decreasing fields, respectively.
This behavior is in a sharp contrast to the magnetization process in \Nd\ described in Section~\ref{sec:M_Nd}.

%_________________________________________________________________________________________________
\subsection{Magnetocaloric effect measurements}	\label{sec:MCE_Pr}
\begin{figure}[tb]
\includegraphics[width=0.95\columnwidth]{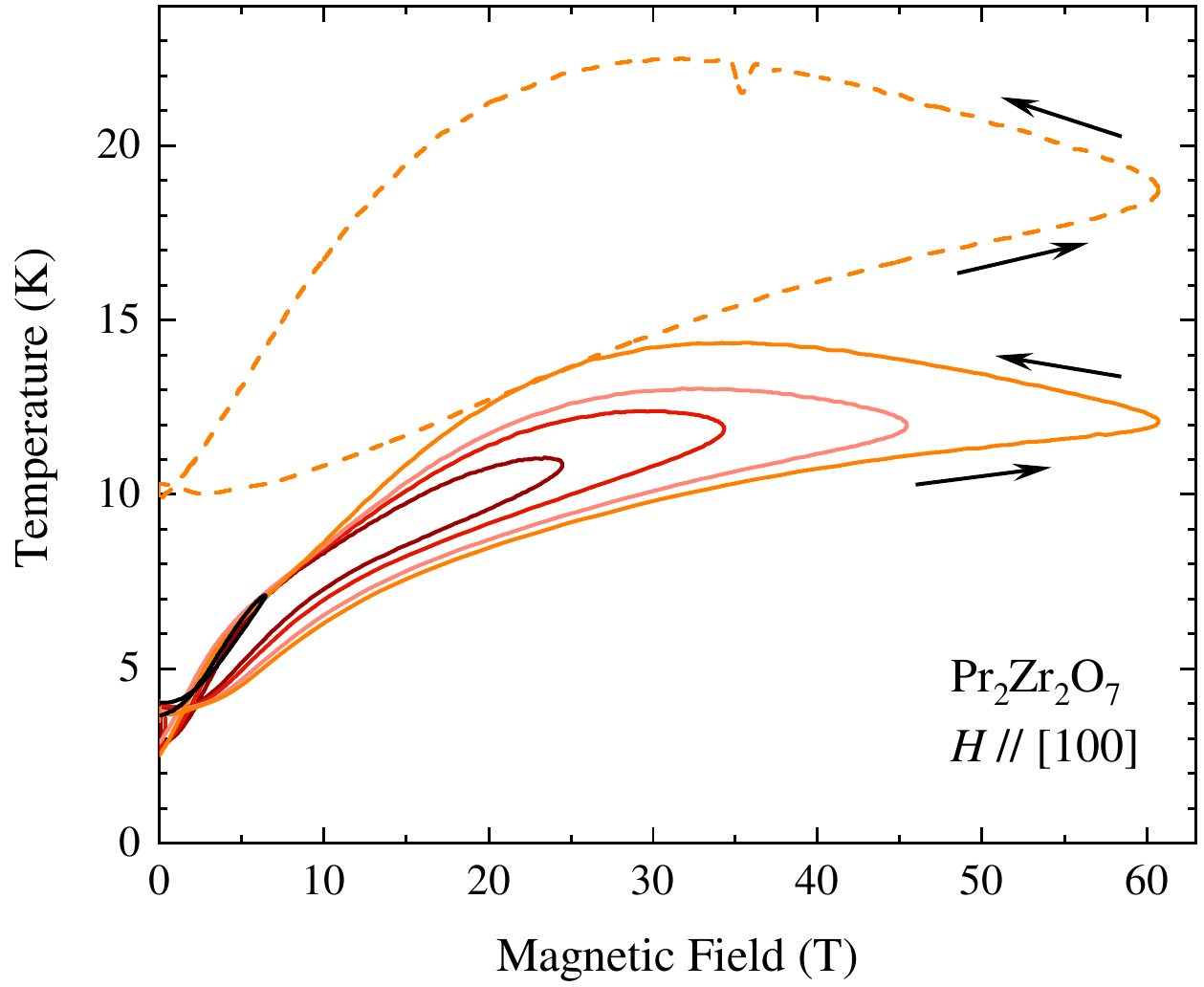}
\caption{Variation of the sample temperature due to the MCE in \Pra\ for the initial temperature of 4~K (solid lines) and 10~K (dotted line).
The colors differentiate the $T(H)$ curves obtained by pulsing the applied field to different maximum values.}
\label{fig:Pr_MCE}
\end{figure}

Figure~\ref{fig:Pr_MCE} shows the MCE data for \HIOO\ for \Pra.
The maximum fields used in the measurements were 6, 25, 34, 45, and 61 T.
The sample's temperature increases with increasing field for all values of the maximum field.
When the applied magnetic field starts to decrease, the sample temperature continues to increase, which is particularly pronounced for field pulses above 30~T.
The MCE in \Pra\ is rather strong -- for an initial temperatures of 4 and 10~K, the highest temperatures reached during a 60~T field pulse are 14 and 23~K, respectively.

\begin{figure}[tb]
\includegraphics[width=0.95\columnwidth]{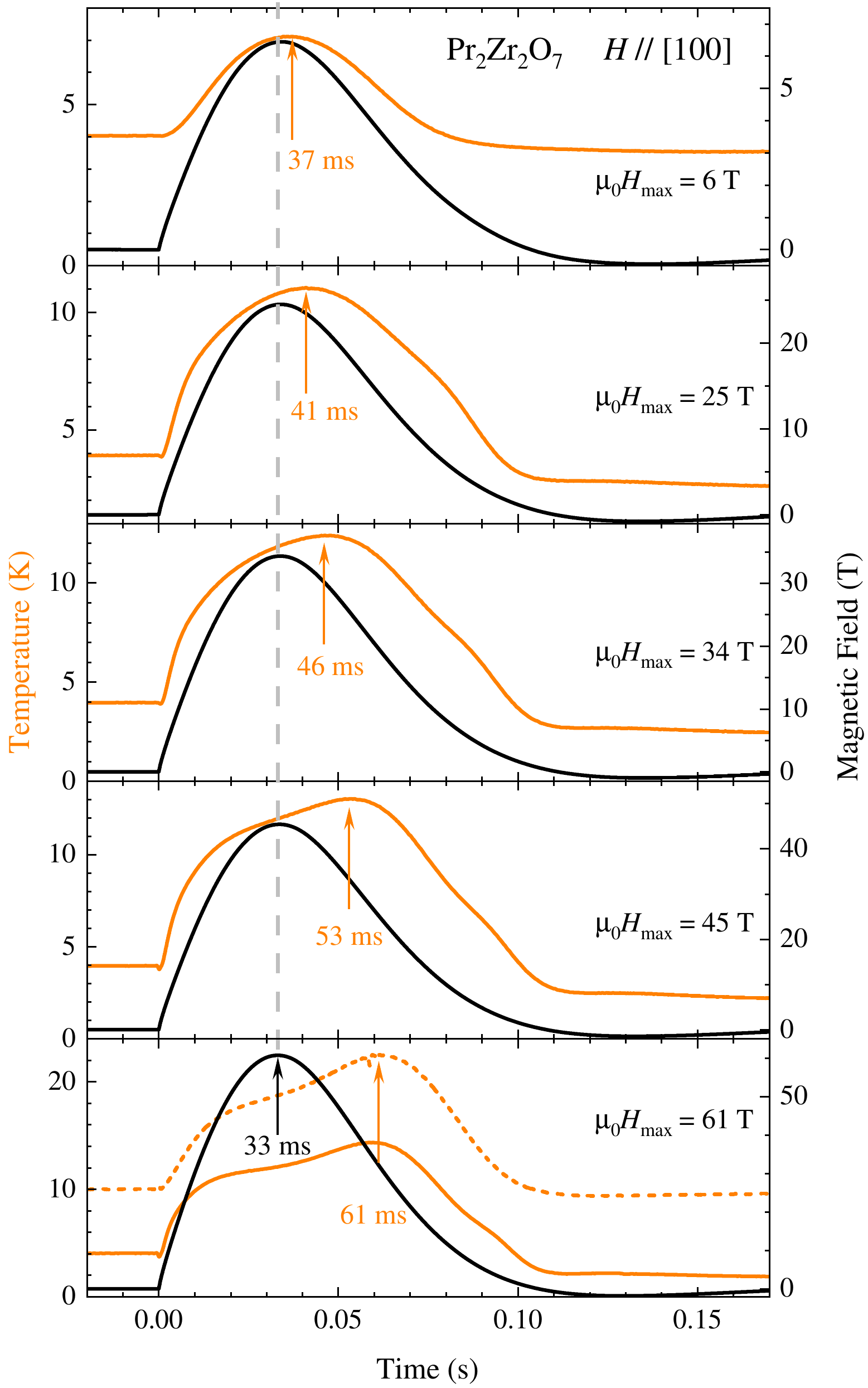}
\caption{Time dependence of the sample temperature of \Pra\ measured while pulsing magnetic field applied along the [110] direction.
The sample temperature and the applied field are presented on the left and right axes, respectively.
The orange arrows indicate the times when the maximum $T$ is recorded, while the field maximum is reached at 33~ms for any value of $H_{\rm max}$.
The initial sample temperature is 4~K (solid orange lines) and 10~K (dashed orange line).}
\label{fig:Pr_MCE_time}
\end{figure}

The MCE data for \Pra\ can also be presented while following the sample temperature as a function of time during the field pulse, see Fig.~\ref{fig:Pr_MCE_time}.
For the lowest $H_{\rm max}$ shown, and therefore the lowest field-sweep rate, the maxima in $H$ and $T$ nearly overlap.
For increasing $H_{\rm max}$, the maximum temperature is reached 8 to 28~ms after the maximum field.
Moreover, increasing $H_{\rm max}$ also results in a significant change in the shape of the $T(t)$ curves, as, instead of a slightly asymmetric single peak mimicking the field pulse, they become more structured.
For $\mu_0 H_{max}=61$~T a weak plateau-like feature is observed in $T(t)$.
A clearer plateau feature is observed in \Nd\ (see Sec.~\ref{sec:MCE_Nd}).

Changing the initial sample temperature from 4 to 10~K does not have much effect on the shape of the $T(t)$ curves.
Given that typical exchange interactions in these materials are in the range $J \sim 1$~K, this observation tends to support the assumption that the MCE is primarily controlled by single-ion effects.

%_________________________________________________________________________________________________
\section{Experimental Results: \Nd} \label{sec:Nd} 
%_________________________________________________________________________________________________
\subsection{Magnetization measurements}		\label{sec:M_Nd}

The conventional approach for converting high-field magnetization data $M(H)$ into absolute units is to measure the same sample in steady fields using vibrating sample or SQUID magnetometers.
As the shapes of the $M(H)$ curves of \Nd\ in pulsed and steady fields differ substantially, a direct comparison between them is not straightforward.
Even at higher temperatures, at which the hysteresis effects in the $M(H)$ curves are less pronounced, the sample temperature changes significantly during the field pulse, leaving a large degree of uncertainty.
We therefore incorporated an additional method for the conversion, which relies on the presumption that, for a particular applied field, the magnetization of the sample depends only on the temperature and, thus, in any field $H$, $M(H)$ cannot exceed the value $M(H,T=0)$.
The $M(H,T=0)$ values could be obtained from steady-field $M(H)$ measurements performed at different temperatures by approximating them to $T=0$~K.
Alternatively, the magnetization results obtained at $T=90$~mK~\cite{Lhotel_2015} could be declared effectively as zero-temperature results.

\begin{figure}[tb]
\includegraphics[width=0.95\columnwidth]{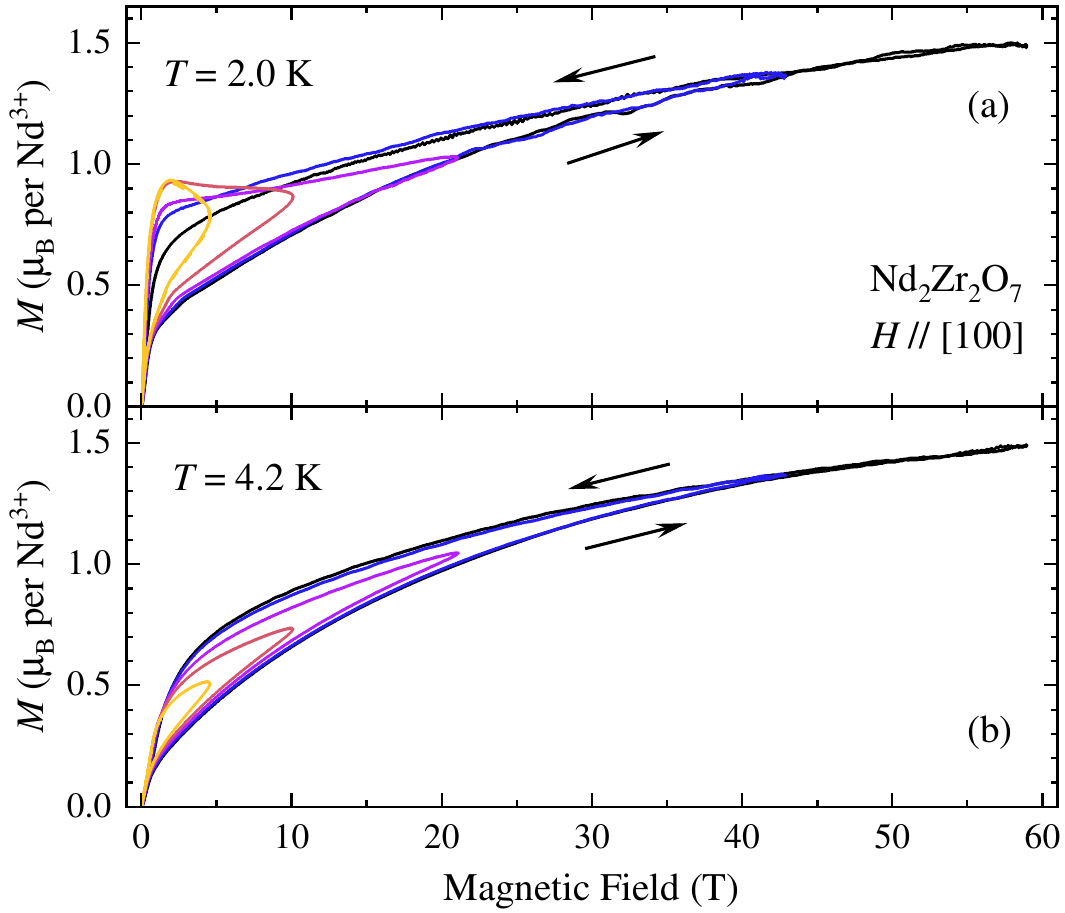}
\caption{Field dependence of the magnetization of \Nd\ for \HIOO\ measured at (a) $T=2.0$~K and (b) $T=4.2$~K.
The colors of the magnetization curves distinguish data collected for magnetic fields pulsed to different maximum values.}
\label{fig:Nd_MH_2_4K}
\end{figure}

Two field directions have been studied in detail, \HIOO\ and \HIIO.
Figure~\ref{fig:Nd_MH_2_4K} shows the results of the magnetization vs field measurements for \HIOO\ performed to various maximum fields for a \Nd\ with the initial temperature of $T=2.0$~K [Fig.~\ref{fig:Nd_MH_2_4K}(a)] and 4.2~K [Fig.~\ref{fig:Nd_MH_2_4K}(b)].
As the field-sweep rate, $dH/dt$, seems to be important -- a large hysteresis is direct proof of this -- and the time-width of the field pulse remains practically constant, we probed the influence of the sweep rate by varying the maximum applied field, $H_{\rm max}$.
The same magnetization curves were reproduced across multiple runs of a given pulse.

There are a few key observations to make while comparing the magnetization data at 2.0 and 4.2~K.
At $T=2.0$~K, the system seems to be slow in reacting to the applied field.
Therefore, for sufficiently high fields, the $M(H)$ curves for different $\mu_0 H_{\rm max}$ (20, 40, 60~T) are identical for increasing field.
For $\mu_0 H_{\rm max}$ of 5 and 10~T (and therefore much lower $dH/dt$), the system is able to react faster to the applied field and the magnetization is considerably higher.
For decreasing fields, the situation is even more dramatic.
The results for $\mu_0 H_{\rm max}$ of 40 and 60~T almost coincide, but for $\mu_0 H_{\rm max}=20$~T a new tendency develops.
The decrease in magnetization with decreasing field is small: the magnetization changes only slightly between 20 and 2~T.
For $\mu_0 H_{\rm max}$ of 5 and 10~T, the $M(H)$ curves are highly unusual, as the magnetization continues to increase in decreasing fields.

The magnetization data taken at 4.2~K also demonstrate a pronounced hysteresis, but they are more conventional in shape, as the magnetization is systematically increasing for increasing field and decreasing  for decreasing field.
However, hysteresis and effects due to field sweep rate are also evident at this higher temperature.

\begin{figure}[tb]
\includegraphics[width=0.95\columnwidth]{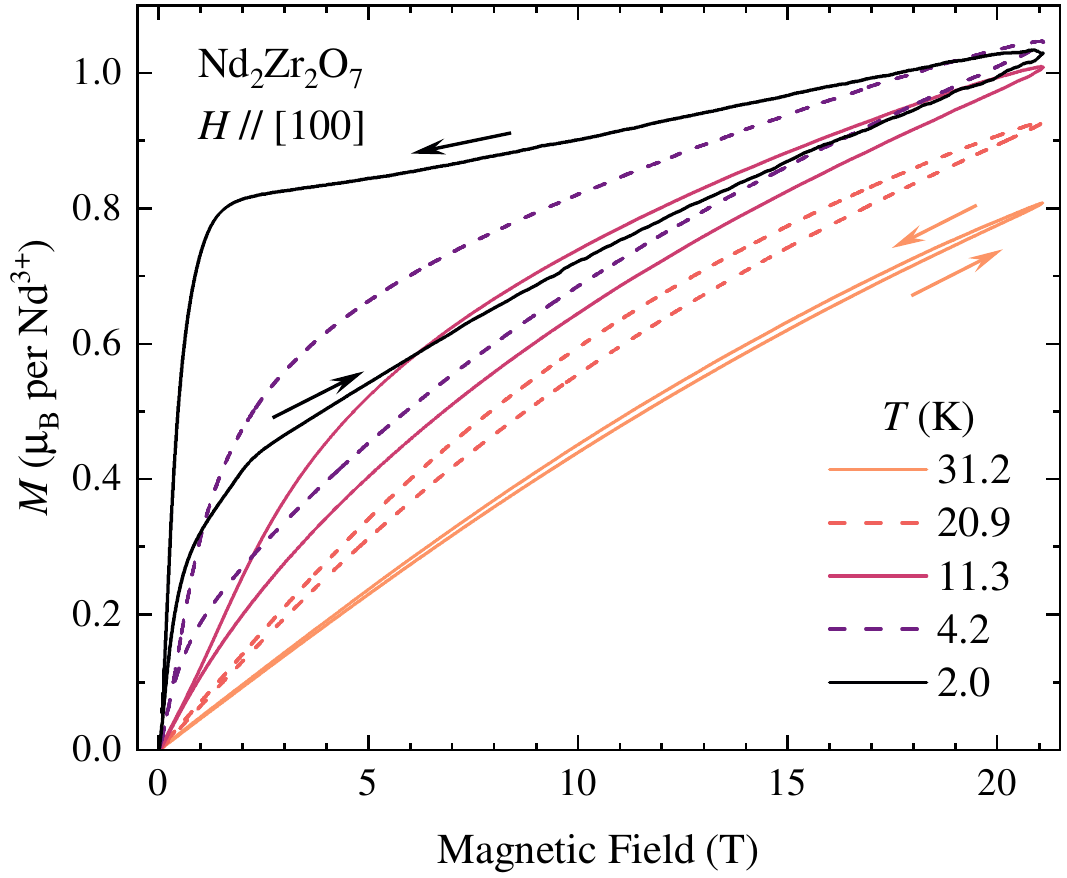}
\caption{Field dependence of the magnetization of \Nd\ for \HIOO\ measured at different temperatures.
For all the curves shown, the magnetization is higher for decreasing applied fields with the hysteresis most pronounced at the lowest temperatures.}
\label{fig:Nd_MH_21T}
\end{figure}

The temperature evolution of the magnetization process is illustrated in Fig.~\ref{fig:Nd_MH_21T}, where we show the data for $\mu_0H_{\rm max}=20$~T for different temperatures.
At the highest temperature studied, $T=31.2$~K, the magnetization behaves conventionally, with only minor hysteresis.
On lowering the temperature, the hysteresis becomes increasingly more pronounced, and the shape of the $M(H)$ curves becomes less conventional, as, at a given field, the magnetization could be twice as high for $H$ decreasing compared to $H$ increasing.
At the lowest temperature shown, $T=2$~K, the magnetization curve is the most unusual, as the decrease in $M(H)$ in decreasing field is small.
The magnetization drops by only ~20\% between 20 and 2~T.
A possible explanation for this effect is that in decreasing fields, the sample temperature also decreases, which therefore results in a higher magnetization.

\begin{figure}[b]
\includegraphics[width=0.95\columnwidth]{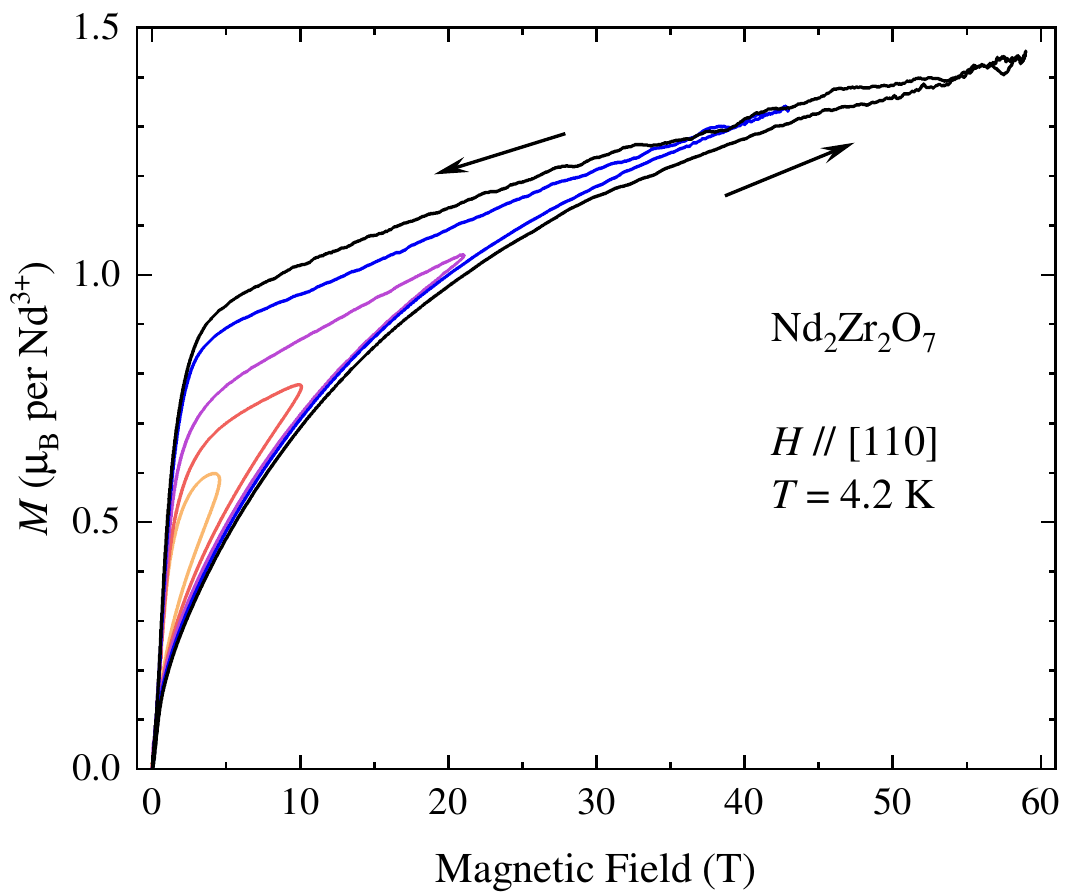}
\caption{Field dependence of the magnetization of \Nd\ measured for \HIIO\ at $T=4.2$~K.}
\label{fig:Nd_MH_4K_110}
\end{figure}

A pronounced hysteresis is also observed for fields in the $[110]$ direction, as shown in Fig.~\ref{fig:Nd_MH_4K_110}.
The shape of the hysteresis is, however, somewhat different to that observed for field aligned along $[100]$.

%_________________________________________________________________________________________________
\subsection{Magnetocaloric effect measurements}	\label{sec:MCE_Nd}
\begin{figure}[tb]
\includegraphics[width=0.95\columnwidth]{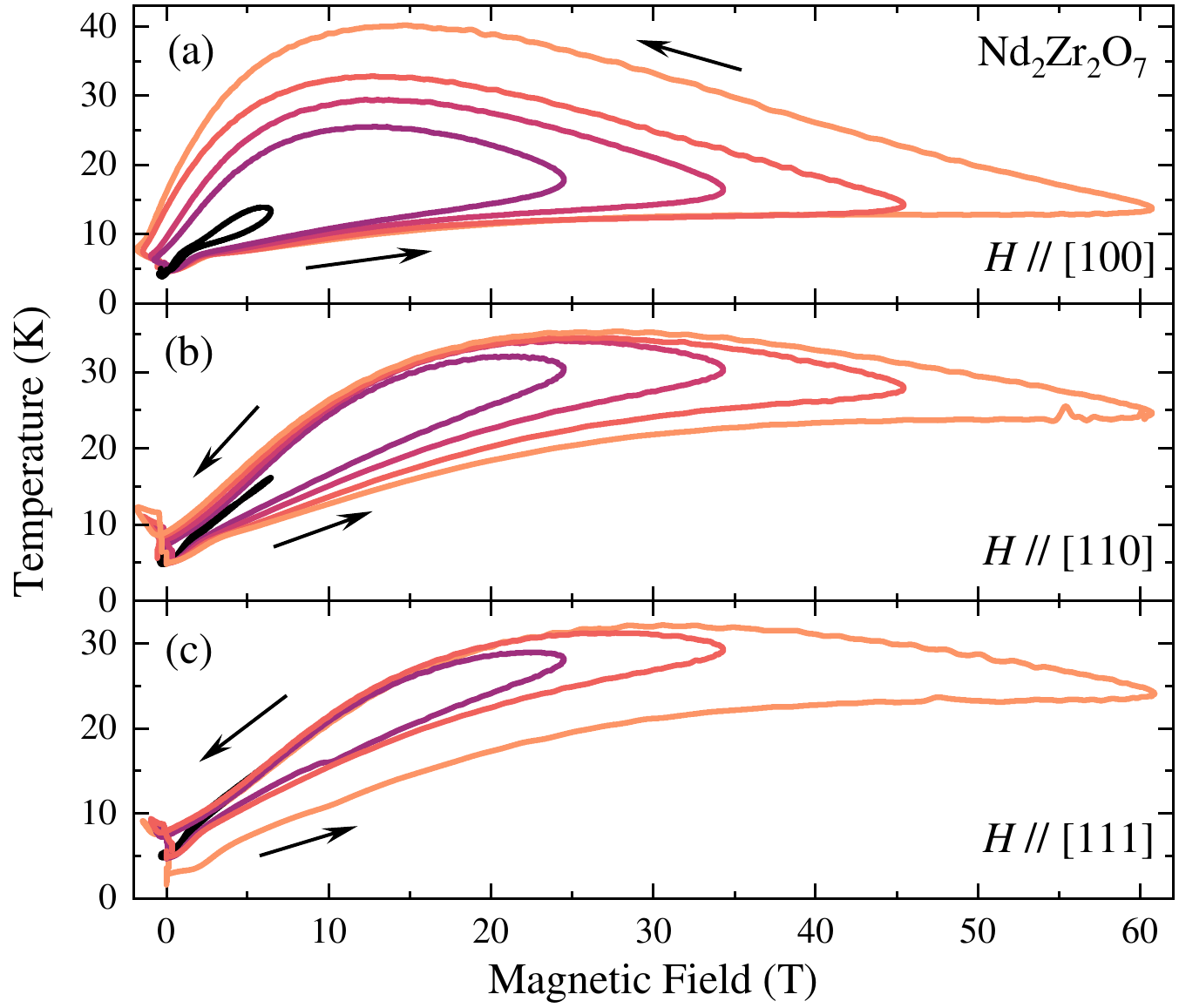}
\caption{Variation of the sample temperature due to the MCE measured in \Nd\ for magnetic fields applied along three high-symmetry directions.
The black arrows indicate the curves for increasing and decreasing magnetic fields.
The colors differentiate the $T(H)$ curves obtained by pulsing the applied field to different maximum values.}
\label{fig:Nd_MCE}
\end{figure}

Figure~\ref{fig:Nd_MCE} shows the MCE data for \HIOO, [110], and [111] for \Nd.
The maximum fields used in the measurements were 7, 25, 35, 45, and 60 T.
Although the MCE curves are anisotropic, their main features look similar for all three field directions.
For the field pulsed to the maximum value of about 60~T, the sample temperature increases with increasing field, the rate of temperature change $dT/dH$ being noticeably lower in higher fields.
When the applied magnetic field starts to decrease, the sample temperature continues to increase, with the rate of temperature change also increasing, particularly for \HIOO.
The sample temperature eventually starts to decrease only when the field is reduced to about half of the maximum value for \HIIO\ and [111] or to about a quarter for \HIOO.

\begin{figure}[tb]
\includegraphics[width=0.95\columnwidth]{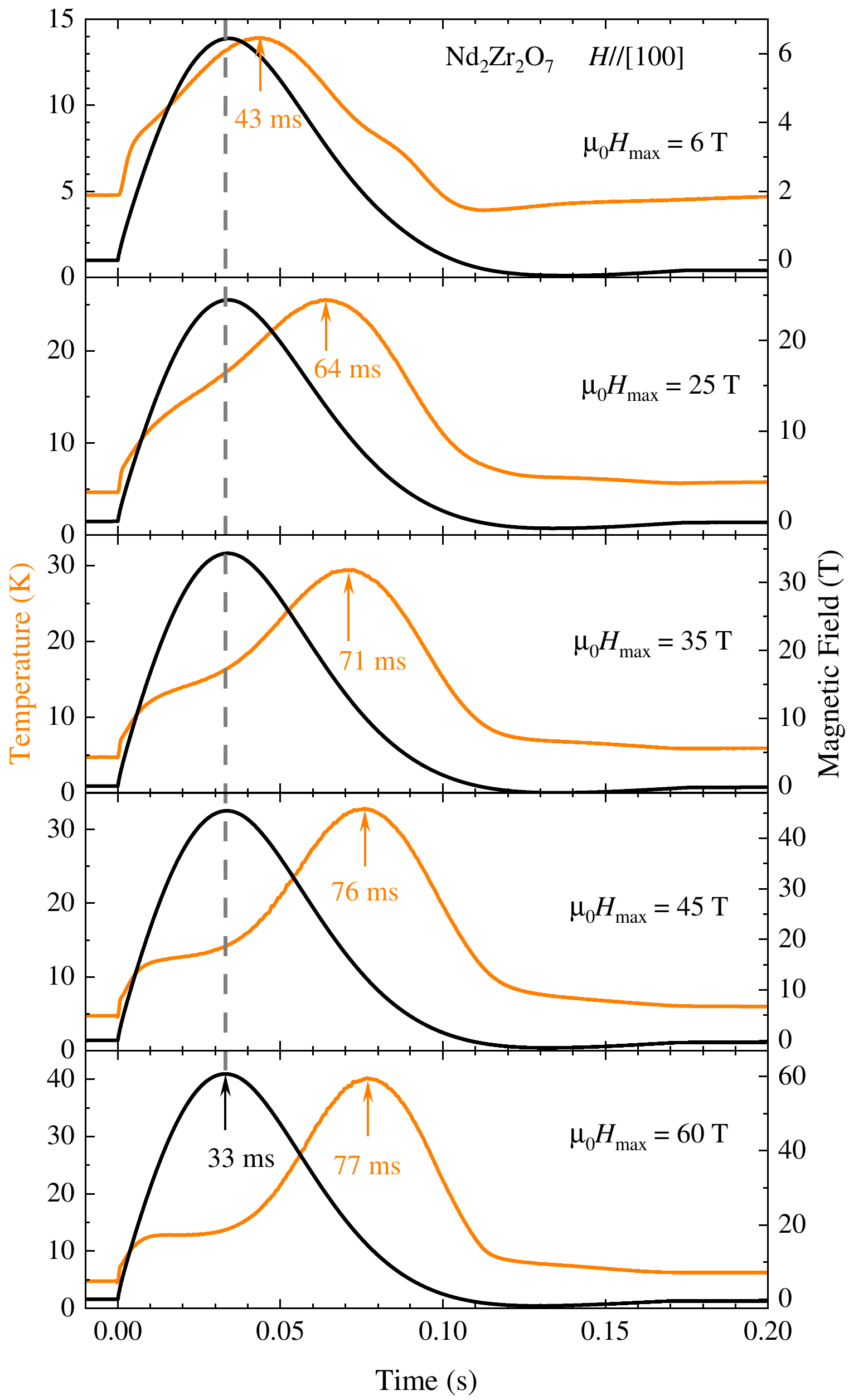}
\caption{Time dependence of the sample temperature of \Nd\ measured while pulsing to different maximum fields $\mu_0 H_{\rm max}$ for \HIOO.
The sample temperature and the applied field are presented on the left and right axes, respectively.
The orange arrows indicate the times when the maximum $T$ is recorded while the field maximum is reached at 33~ms for any value of $H_{\rm max}$.}
\label{fig:Nd_MCE_time}
\end{figure}

Instead of plotting the field dependence $T(H)$, it is enlightening  to show the same MCE data as a function of time, as presented in Fig.~\ref{fig:Nd_MCE_time} for \HIOO.
This presentation allows one to check how quickly the sample temperature tracks the variations in applied fields.
The duration of the magnetic field pulse (shown as black curves) remains practically identical for all values of the maximum field, with the maximum reached around 33~ms after the start of the pulse.
The maximum of the sample temperature (represented by orange curves) is reached with a significant time lag, ranging from 10~ms for $\mu_0 H_{\rm max}=6$~T to about 34~ms for the ten times higher maximum field.
Such a long time lag is unusual; its value is several orders of magnitude higher than what is typically observed in magnetic materials exposed to pulsed fields.

The shape of the $T(t)$ curves is also unusual, particularly for higher applied fields (and therefore higher rates of field variation, $dH/dt$).
For all fields, the initial increase in the sample temperature (straight after the start of a field pulse) is rather fast.
After approximately one millisecond, the rate of increase in temperature slows down.
The higher the $dH/dt$, the more pronounced this effect is.
For the maximum applied field of 60~T, sample temperature remains stable at $T \approx 13$~K between 11 and 30~ms, while the applied field almost doubles in value.
We discuss these results in more detail in the context of a phenomenological model in Sec.~\ref{sec:theory}.

%_________________________________________________________________________________________________
\section{Theoretical modeling}		\label{sec:theory}

In order to model the observed magnetization $M(H)$ and MCE $T(H)$ curves we begin with some assumptions, motivated by the data.
\begin{enumerate}[label=\roman*.]
\item{The response of the materials to the magnetic field is mostly controlled by single-ion physics (crystal electric fields) rather than exchange interactions.
This is supported by the observation that the  hysteresis appears for starting temperatures $T_0 \gtrsim 10$~K [Figs.~\ref{fig:Pr_MCE}, \ref{fig:Nd_MCE}] whereas exchange interactions in these materials are $\sim 1$~K.}
\item The magnetic and lattice DOFs are out of thermal equilibrium on the timescale of the field pulse.
This is evident from the discrepancy between the magnetization $M(H)$ curves -- which imply higher effective temperatures on the up-sweep -- and the direct temperature measurements, which are higher on the down-sweep.
The direct temperature measurements reflect the phonon temperature \( T_{\mathrm{ph}} \), which we distinguish from the magnetic subsystem's effective temperature \( T_{\mathrm{CEF}} \). The idea of distinct effective temperatures for magnetic and lattice degrees of freedom has a long history \cite{Sanders_1977} and is relevant to a range of experiments on magnetic materials \cite{Liao_2014, Li_arxiv}.
\item The system's non-equilibrium behavior and its dependence on the field-sweep rate indicate that any realistic description of the MCE must be dynamical (i.e. time dependent).
\item{Over timescales shorter than the rate of heat transfer between CEF subsystem and phonons the magnetic DOFs should respond adiabatically to the applied field.}
\end{enumerate}

Based on the above assumptions, we propose the following coupled differential equations as a model for the evolution of the temperature as a function of time:
\begin{widetext}
\begin{eqnarray}
&&\frac{d T_{\rm CEF}}{dt}=- \frac{T_{\rm CEF}}{C_{\rm CEF}({\bf H},T)} \left( \frac{\partial S}{\partial H} \right)_T \frac{dH}{dt}
- \frac{\lambda_1 k_{\rm B}}{C_{\rm CEF} ({\bf H},T)} (T_{\rm CEF}-T_{\rm ph}) 
\label{eq:cef_time}
\\
&&\frac{d T_{\rm ph}}{dt}= \frac{\lambda_1 k_{\rm B}}{C_{\rm ph} (T)} (T_{\rm CEF}-T_{\rm ph})-\frac{\lambda_2 k_{\rm B}}{C_{\rm ph}(T)} (T_{\rm ph}-T_f).
\label{eq:phonontime}
\end{eqnarray}
\end{widetext}

The first term in Eq.~(\ref{eq:cef_time}) describes the adiabatic magnetocaloric effect as applied to the CEF subsystem.
This subsystem is then coupled to the lattice by an ({\it a priori} unknown) thermal coupling $\lambda_1$.
The lattice is then coupled to the external environment (thermal reservoir), which is kept at the fixed temperature $T_f$, via a thermal coupling $\lambda_2$. 
$\lambda_1$ and $\lambda_2$ have dimensions of inverse time, so should be treated as relaxation rates.
To begin with, we keep $\lambda_1$ and $\lambda_2$ constant with respect to the applied field and temperature, and discuss later the possible dependence of $\lambda_1$ on field strength and sweep rate.

$dH/dt$ is known from the experimental field-pulse profile, and the lattice heat capacity $C_{\rm ph}$ can be estimated from experiments on the nonmagnetic analogue La$_2$Zr$_2$O$_7$~\cite{Xu_2019}.
The crystal-field heat capacity $C_{\rm CEF}(H,T)$ and the isothermal rate of entropy change 
$\left( \frac{\partial S}{\partial H} \right)_T$ can be calculated from modeling the CEF spectrum (see Appendices~\ref{app:therm_cef} and \ref{app:cef_details}).

Eqs.~(\ref{eq:cef_time}) and (\ref{eq:phonontime}) are solved numerically to yield $T_{\rm CEF}(t)$ and $T_{\rm ph}(t)$.
Using the known field-pulse profile $H(t)$, we obtain temperature as a function of magnetic field on the up and down sweep of the field.
Magnetization can also be calculated based on the evolution of the CEF spectrum with field, with the temperature of the magnetic DOFs ($T_{\rm CEF}$) taken into account.

We focus on fields in the $[100]$  direction.
This simplifies the calculations because for this field direction all sites in the unit cell have equivalent single-ion CEF spectra.

\subsection{Application to \Pra}
\label{subsec:model_PZO}

The crystal field spectrum of \Pra\ is modeled using parameters from Ref.~[\onlinecite{Bonville_2016}].
These parameters allow us to calculate the evolution of the CEF levels on each Pr$^{3+}$ site as a function of applied magnetic field.

\begin{figure}[tb]
\includegraphics[width=0.95\columnwidth]{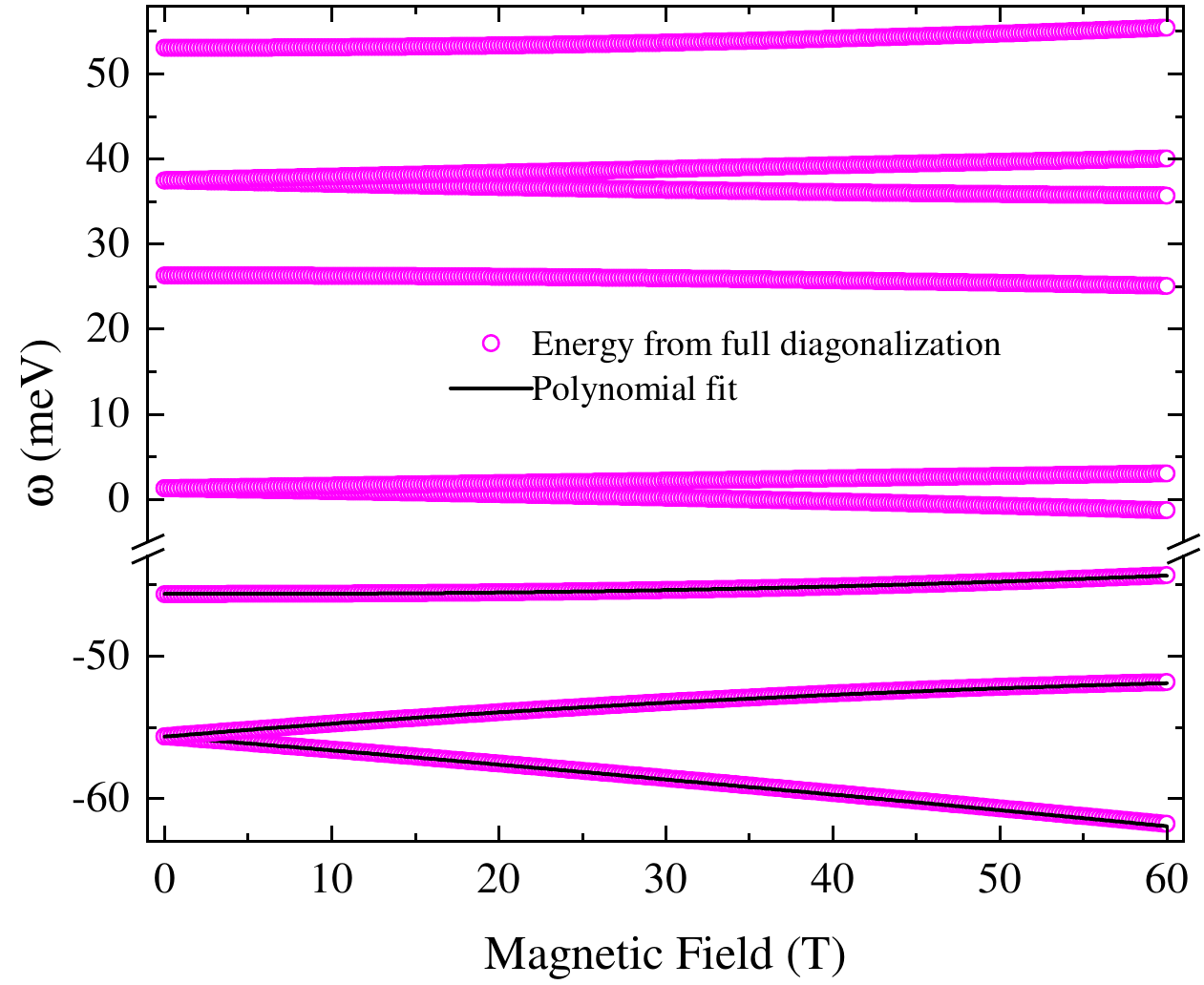}
\caption{Evolution of the CEF spectrum of the Pr$^{3+}$ ions in a [100] field based on the crystal-field parameters given in Ref. [\onlinecite{Bonville_2016}]. 
For the lowest three energy levels, a quadratic fit of their field dependence is also shown. 
This quadratic fit is used in calculations to enable analytic differentiation of the spectrum with respect to field.
}
\label{fig:Pr_spec}
\end{figure}

\begin{figure}[tb]
\includegraphics[width=0.95\columnwidth]{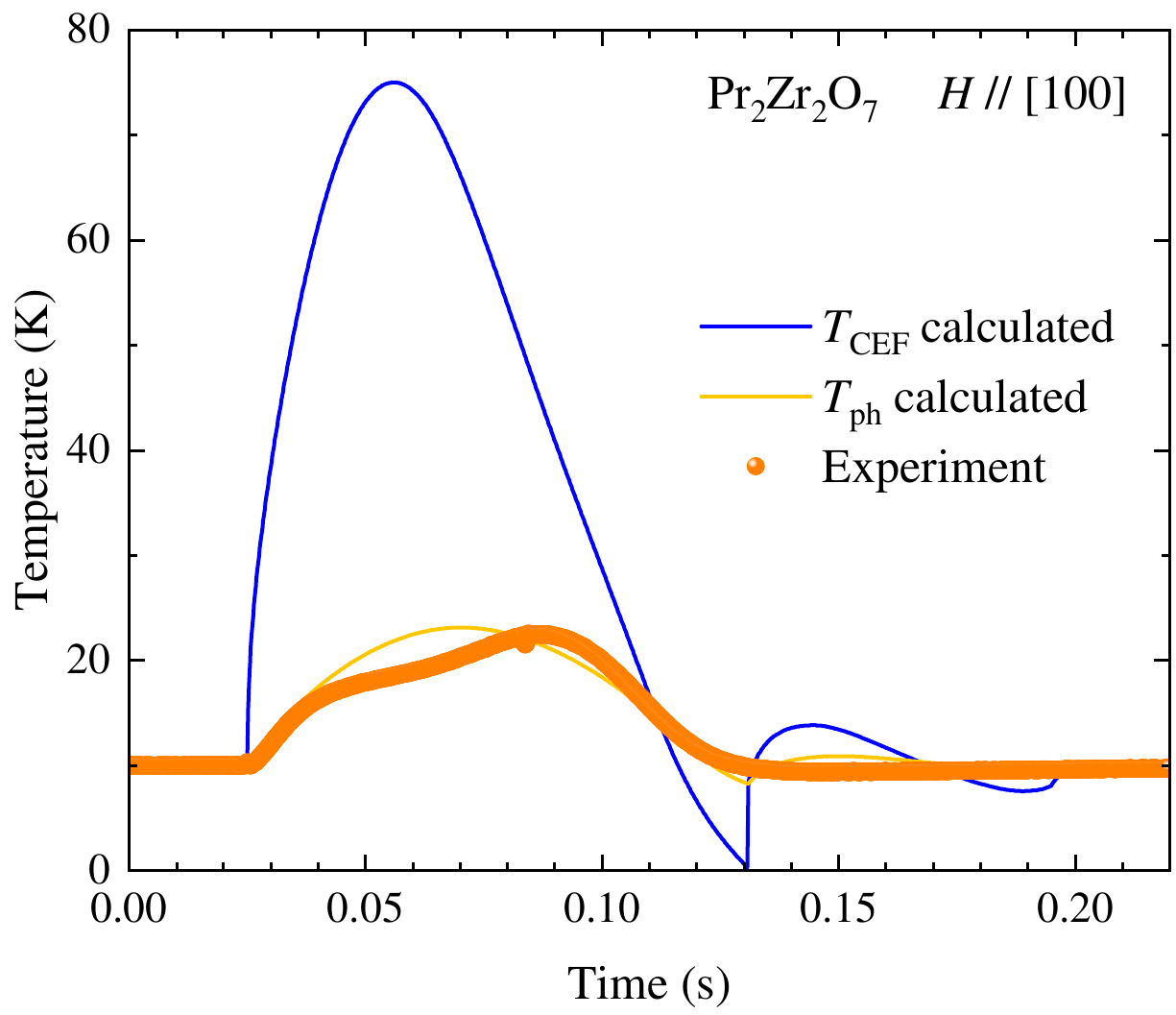}
\caption{Simulated crystal-field temperature $T_{\rm CEF} (t)$ and phonon temperature $T_{\rm ph} (t)$ for a pulsed magnetic field of 60~T and initial temperature $T_0 = 10.1$~K in \Pra, compared with experimental data. Calculations have been made using Eqs.~(\ref{eq:cef_time}) and (\ref{eq:phonontime})
and a pulse profile $H(t)$ reflecting that used in experiments (see Sec.~\ref{sec:experimental}).
The values of the thermal couplings $\lambda_1 = 2.56$~s$^{-1}$ and $\lambda_2 = 8.44$~s$^{-1}$ have been set by optimizing agreement with experimental data using a grid search of parameter space.}
\label{fig:Pr_T(t)}
\end{figure}

\begin{figure}[tb]
\includegraphics[width=0.95\columnwidth]{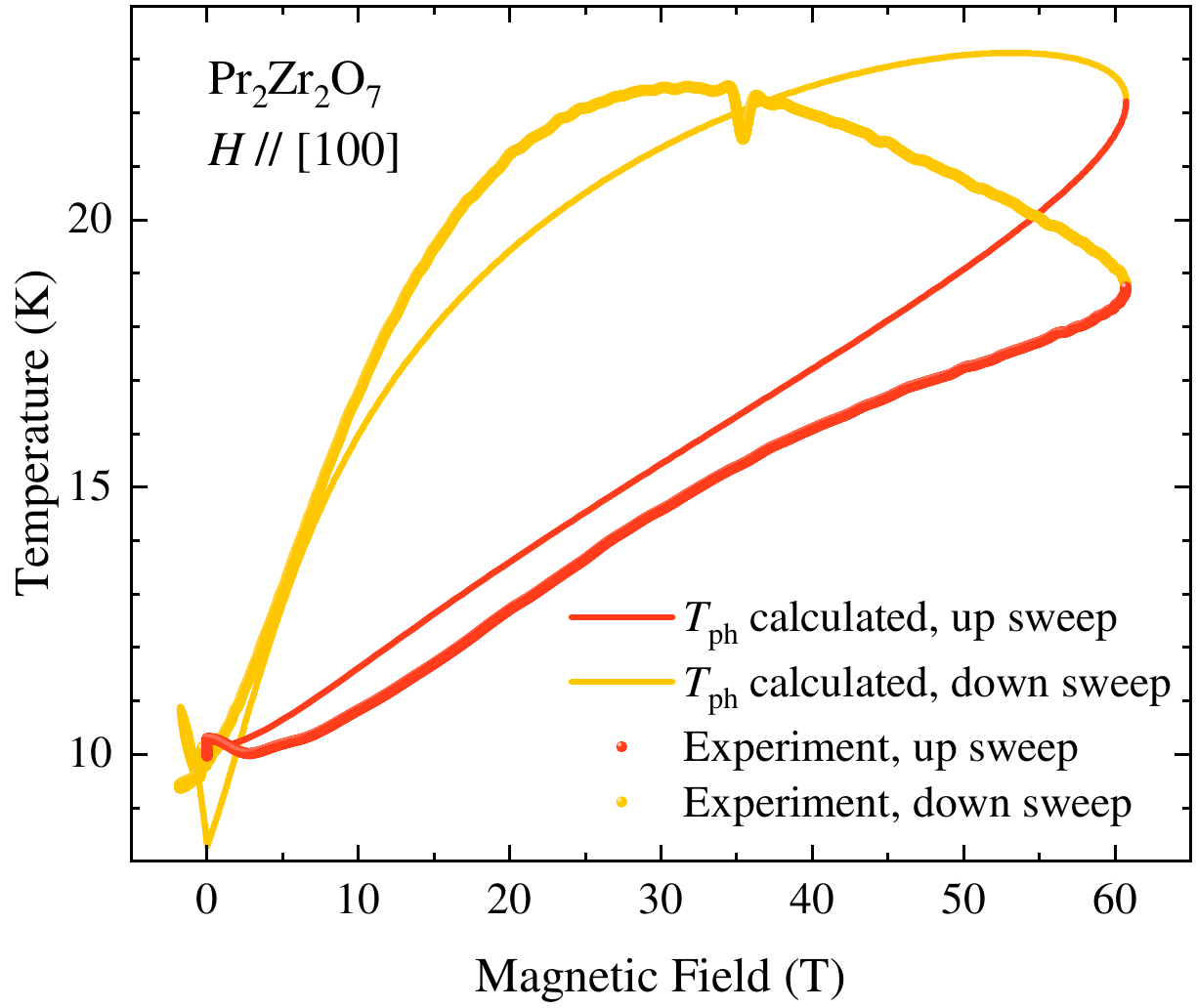}
\caption{Simulated phonon temperature $T_{\rm ph} (H)$ plotted against field strength for a pulsed magnetic field of maximum strength $H_{max} = 60$~T and initial temperature $T_0=10.1$~K in \Pra, compared with experimental data.
Calculations have been made using Eqs.~(\ref{eq:cef_time}) and (\ref{eq:phonontime}) and a pulse profile $H(t)$ reflecting that used in experiments (see Sec.~\ref{sec:experimental}).
The values of the thermal couplings $\lambda_1 = 2.56$~s$^{-1}$ and $\lambda_2 = 8.44 $~s$^{-1}$ have been set by optimizing agreement with experimental data using a grid search of parameter space.}
\label{fig:Pr_T(H)}
\end{figure}

\begin{figure}[tb]
\includegraphics[width=0.95\columnwidth]{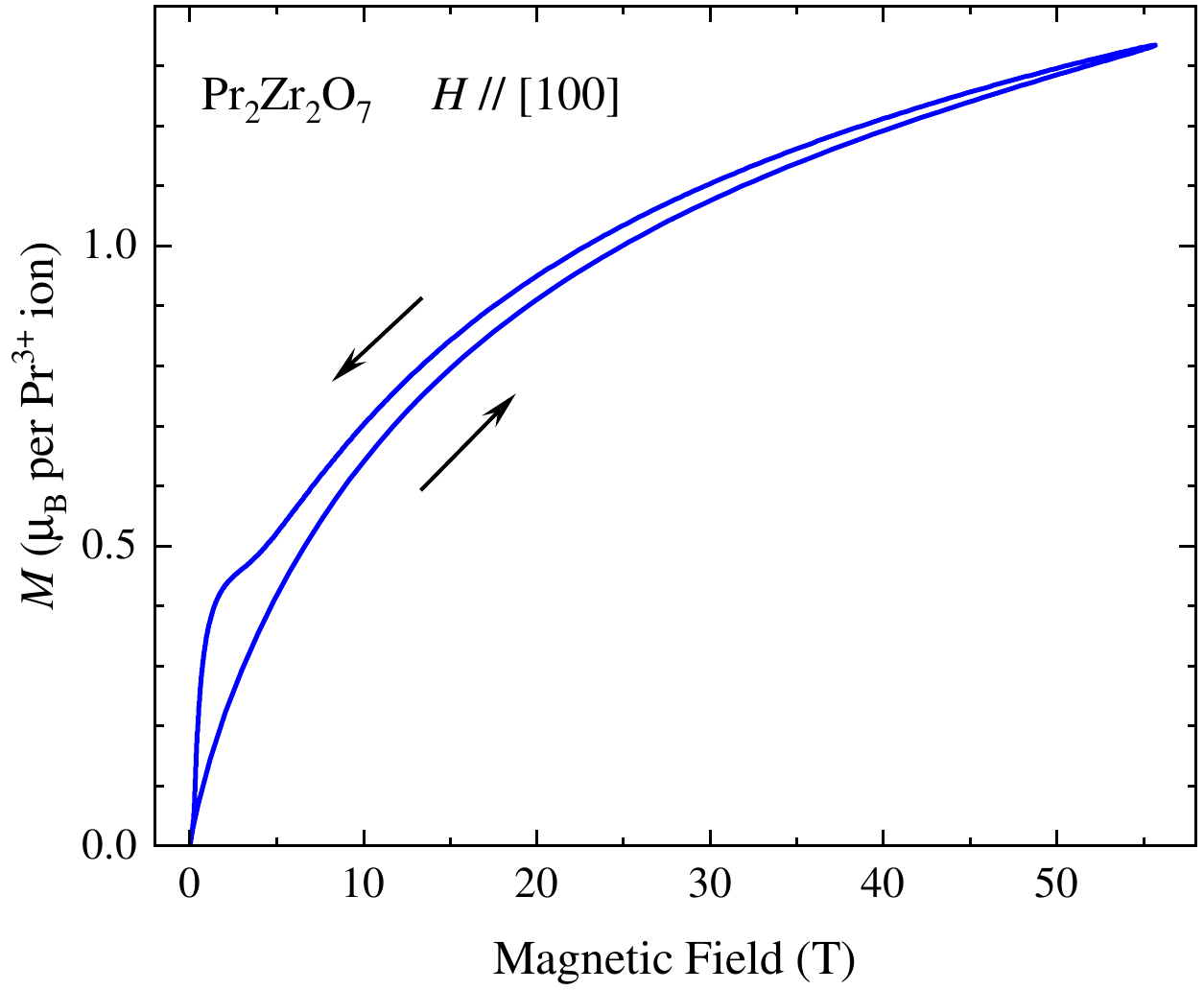}
\caption{Predicted magnetization curve for \Pra\ based on Eqs.~(\ref{eq:cef_time}) and~(\ref{eq:phonontime}) and the parameters given in Eq.~(\ref{eq:lambda_vals_Pr}).
Calculations are made assuming a starting temperature $T=10$~K and a field along the [100] direction. 
This reproduces the relative lack of hysteresis and the failure to reach saturated magnetization even at 60~T seen in the experiments [Fig.~\ref{fig:Pr_MH}].
This calculation uses a faster field pulse $H(t)$ than the calculations of the MCE [Figs.~\ref{fig:Pr_T(t)}-\ref{fig:Pr_T(H)}], to approximate the pulse used in the magnetization measurements (see Sec.~\ref{sec:experimental}.)}
\label{fig:Pr_mag_calc}
\end{figure}

Figure~\ref{fig:Pr_spec} shows the calculated CEF spectrum as a function of applied magnetic field, including all nine states expected for a $J = 4$ ion.
A significant energy gap separates the lowest three states from the remaining six, allowing us to simplify the calculations by retaining only the lowest three levels in our model.

Thermodynamic quantities such as \( \left( \frac{\partial S}{\partial H} \right)_T \) depend on the field derivatives of the CEF eigenenergies \( \omega_n \).
To simplify these calculations each \( \omega_n(H) \) is fitted by a second-order polynomial in field, from which we can then obtain \( \frac{\partial \omega_n}{\partial H} \) analytically.
Figure~\ref{fig:Pr_spec} shows the agreement between the fitted curves and the numerically computed spectrum.

The MCE was calculated for \Pra\ using Eqs. (\ref{eq:cef_time}) and (\ref{eq:phonontime}) and the experimental field-pulse profile (see Sec.~\ref{sec:experimental}).
The resulting time dependence of the CEF and phonon temperatures  and is shown in Fig. \ref{fig:Pr_T(t)}, and compared with the experimental MCE data.
The parameters $\lambda_1$ and $\lambda_2$ are set to:
\begin{eqnarray}
\lambda_1 = 2.56~{\rm s}^{-1}, 
\lambda_2 = 8.44~{\rm s}^{-1}
\label{eq:lambda_vals_Pr}
\end{eqnarray}
having optimized agreement between the calculated $T_{\rm ph}(t)$ and the experimentally measured temperatures using a grid search.

With these parameters, we find good agreement between the experimental MCE and $T_{\rm ph}(t)$, although the weak plateau-like feature before the maximum in the experimental data is not reproduced.
We find large differences between $T_{CEF}(t)$ and $T_{\rm ph} (t)$ over the course of the experiment.

The same data is presented in Fig.~\ref{fig:Pr_T(H)}, but now as a function of applied magnetic field, rather than time, and color-coded to indicate the up and down sweeps of the field.
The model reproduces the magnitude of the observed MCE and the large hysteresis, although not the exact shape of the hysteresis loop.

Fig. \ref{fig:Pr_mag_calc} shows the predicted magnetization curve in a pulsed-field experiment.
For these calculations we took into account the different pulse profile used in the magnetization experiments (see Sec.~\ref{sec:experimental}).
Comparing this with the experimental observations [Fig.~\ref{fig:Pr_MH}], we observe that the theory reproduces the relative lack of hysteresis observed experimentally, as well as the lack of saturation even at 60~T.
This is because the field pulse used in the magnetization experiments is fast compared to that used in the MCE measurements.
With this fast pulse, heat does not have time to transfer from magnetic to lattice degrees of freedom, and the system effectively follows an adiabatic magnetization curve but with the relevant heat capacity being given by $C_{\rm CEF} (T,H)$ rather than the total heat capacity.
The theoretical curve predicts lower values of magnetization than observed experimentally.
This is probably due to the very low value of the calculated $C_{\rm CEF} (T,H)$ at the starting temperature $T=10$~K, which leads to very rapid heating of the magnetic DOFs.
In the real system,  contributions to the heat capacity due to interactions or disorder-induced CEF splitting would likely smooth out this rapid heating and lead to a higher magnetization.

Despite the limitations described above, the model given in Eqs.~(\ref{eq:cef_time}) and~(\ref{eq:phonontime}) gives a reasonable account for the observations on \Pra\ - namely the substantial hysteresis in the MCE, the overall magnitude of the MCE, and the small hysteresis in the magnetization curve. 

\subsection{Application to \Nd}        \label{subsec:model_NZO}

\begin{figure}[b]
\includegraphics[width=0.95\columnwidth]{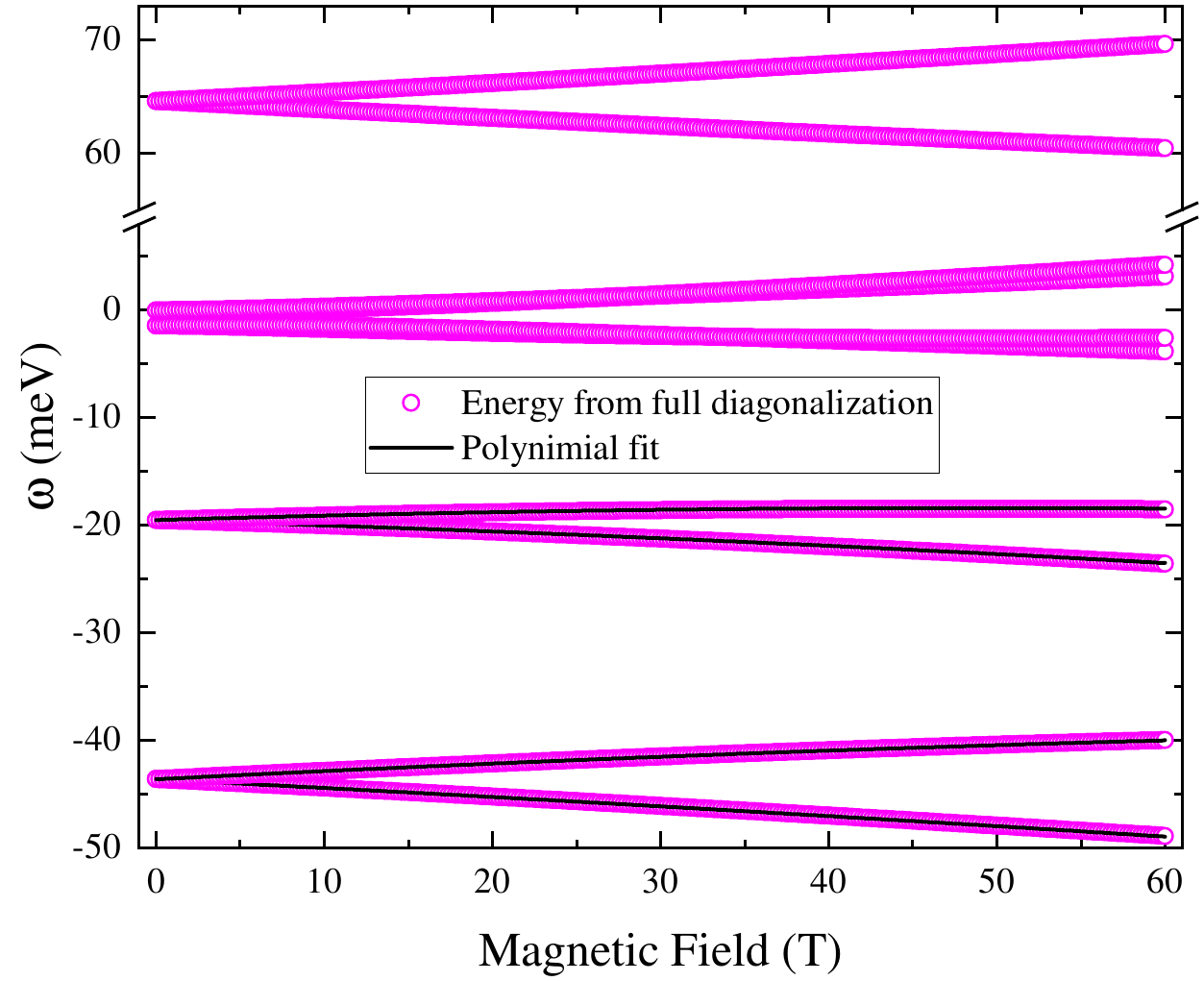}
\caption{Evolution of the CEF spectrum of the Nd$^{3+}$ ions for fields aligned along [100] based on the crystal-field parameters given in Ref.~[\onlinecite{Xu_2015}]. 
For the lowest four energy levels, a quadratic fit of their field dependence is also shown. 
This quadratic fit is used in calculations to enable analytic differentiation of the spectrum with respect to field.}
\label{fig:Nd_spec}
\end{figure}

To model the behavior of \Nd, we used the crystal-field parameters from Ref.~[\onlinecite{Xu_2015}].
The predicted evolution of the CEF spectrum with applied field is shown in Fig.~\ref{fig:Nd_spec}.
In zero applied field, there are 5 Kramers doublets, making up the $2J+1=10$ states expected for a $J=9/2$ ion.
These are split by the magnetic field, and the predicted evolution with field is close to linear all the way up to 60~T. 

As with \Pra, we simplify the description by including only the lowest pair of zero-field multiplets in the model (this time comprising 4 instead of 3 states) and employ a quadratic fit of their energies with field to enable analytic differentiation (see  Fig.~\ref{fig:Nd_spec}).

\begin{figure}[tb]
\includegraphics[width=0.95\columnwidth]{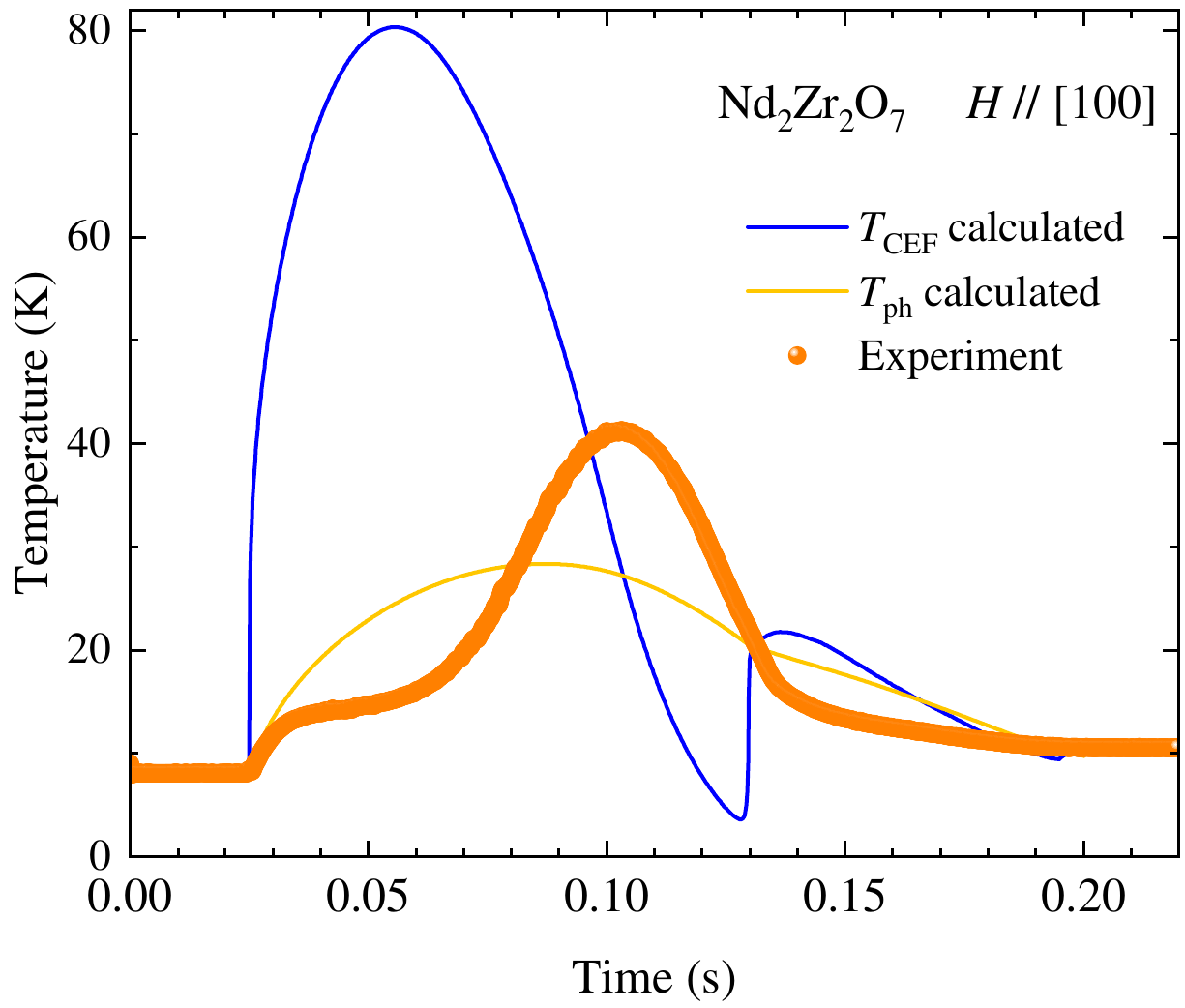}
\caption{Simulated crystal-field temperature
$T_{\rm CEF} (t)$ and phonon temperature $T_{\rm ph} (t)$ for a pulsed magnetic field of maximum strength $\mu_0 H_{max}=60$~T and initial temperature $T_0=9$~K in \Nd, compared with experimental data. Calculations have been made using Eqs.~(\ref{eq:cef_time}) and (\ref{eq:phonontime}) and a pulse profile $H(t)$ reflecting that used in experiments (see Sec.~\ref{sec:experimental}).
The values of the thermal couplings $\lambda_1=1.9$~s$^{-1}$ and $\lambda_2 = 2.5 $~s$^{-1}$ have been set by optimizing agreement with experimental data using a grid search of parameter space.}
\label{fig:Nd_T(t)}
\end{figure}

\begin{figure}[tb]
\includegraphics[width=0.95\columnwidth]{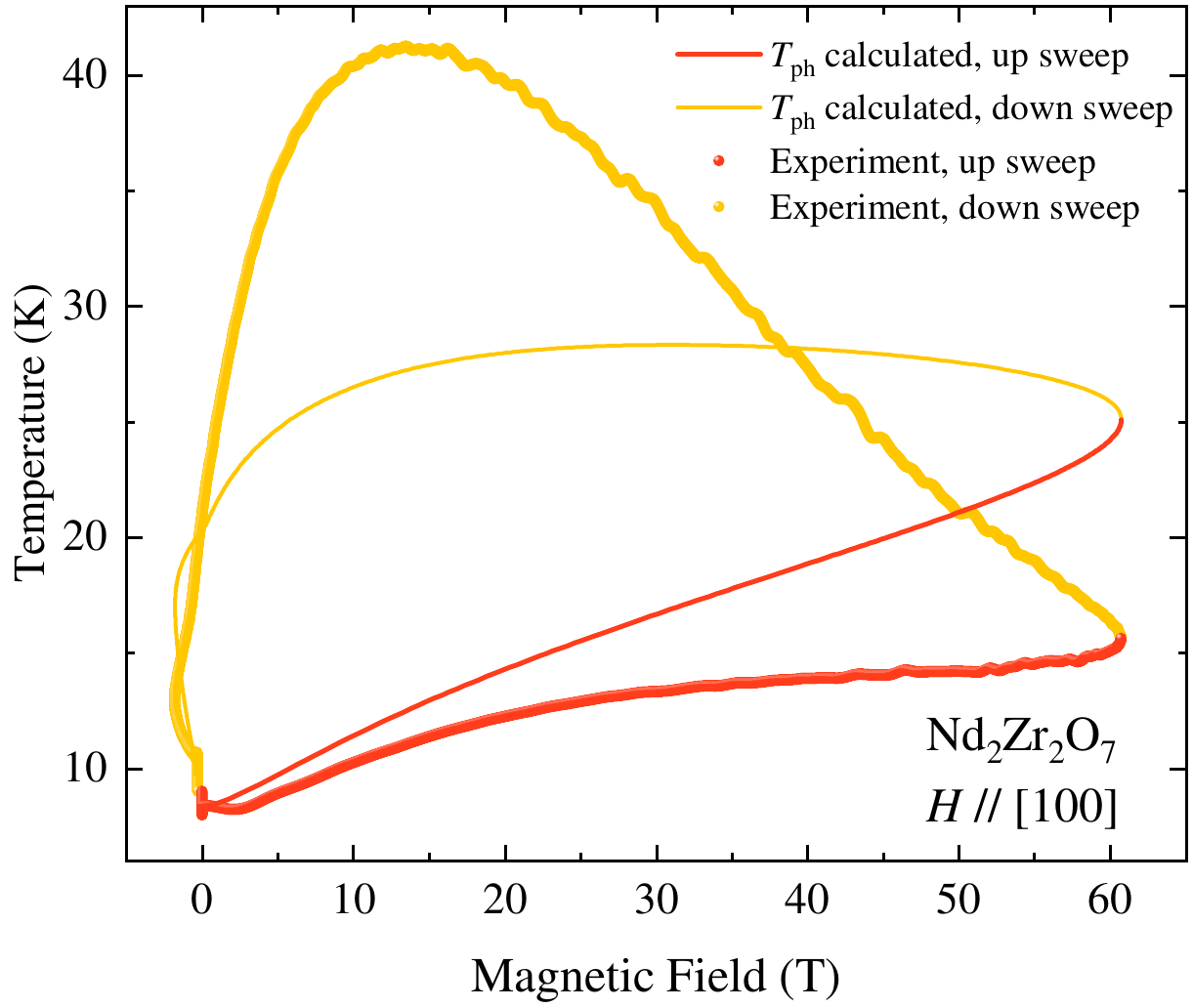}
\caption{Simulated phonon temperature plotted against field strength for a pulsed magnetic field of 60~T and initial temperature $T_0=9$~K in \Nd, compared with experimental data.
Calculations have been made using Eqs.~(\ref{eq:cef_time}) and (\ref{eq:phonontime}) and a pulse profile $H(t)$ reflecting that used in experiments (see Sec.~\ref{sec:experimental}).
The values of the thermal couplings $\lambda_1 =1.9$~s$^{-1}$ and $\lambda_2=2.5$~s$^{-1}$ have been set by optimizing agreement with experimental data using a grid search of parameter space.}
\label{fig:Nd_T(H)}
\end{figure}

The MCE in \Nd\ was simulated using Eqs.~(\ref{eq:cef_time}) and~(\ref{eq:phonontime}).
A grid search of parameter space was used to identify optimal parameters $\lambda_1$ and $\lambda_2$.
The best agreement was obtained with
$$
\lambda_1 = 1.9~{\rm s}^{-1} \ \ \ \ 
\lambda_2 = 2.5~{\rm s}^{-1}.
$$
The resulting MCE curve is shown as a function of time in  Fig.~\ref{fig:Nd_T(t)}
and as a function applied field in Fig.~\ref{fig:Nd_T(H)}.

While the theory captures some features of the data - particularly the presence of the anomalously large hysteresis loop - the agreement is evidently much worse than with the \Pra\ data in Fig.~\ref{fig:Pr_T(H)}.
In particular, the presence of the plateau in $T(t)$ is not captured by the theory.
We were unable to find any parameter set which would reproduce this behavior based on 
Eqs. (\ref{eq:cef_time})-(\ref{eq:phonontime}) with constant relaxation rates $\lambda_1$ and $\lambda_2$.
We therefore seek to refine the model in the following section.

\subsection{Refinement to the model: field and sweep-rate dependence of $\lambda_1$}      \label{subsec:sweep_rate}

In order to improve the model, we need to move beyond the assumption of constant relaxation rates $\lambda_1$ and $\lambda_2$.
While the simplest refinements would be to allow explicit dependence on $T$ or $H$, the experimental data in Fig.~\ref{fig:Nd_MCE_time} suggest that this will be insufficient.

The plateau in $T(t)$, which is evident in the data for $\mu_0 H_{max}=35$, 45, and 60~T, ends at a time closely coinciding with the maximum applied field. 
This is true, even though both the maximum field  and the temperature at maximum field are different.
Thus, the rise in relaxation rate that signals the end of the plateau is not associated to reaching any particular value of $T$ or $H$. Instead, it is associated to the transition from the up to the down of the field.

We can distinguish the up sweep from the down sweep using the quantity:
$$
H \frac{dH}{dt} = \frac{1}{2} \frac{d (H^2)}{dt} \,
$$
which is proportional to the rate of change of energy stored in the magnetic field.

Intuitively, the relaxation rates cannot be negative and also cannot be arbitrarily large. We therefore introduce a dependence of  $\lambda_1$  on  $H \frac{dH}{dt}$, which interpolates between two finite, positive, values $\lambda_{1\uparrow}$ for large positive $H \frac{dH}{dt}$ (up sweep)  and $\lambda_{1\downarrow}$ for large negative $H \frac{dH}{dt}$ (down sweep):
\begin{eqnarray}
\lambda_1 = \frac{1}{2} (\lambda_{1 \uparrow} + \lambda_{1 \downarrow}) + \frac{1}{2} (\lambda_{1 \uparrow} - \lambda_{1 \downarrow}) \tanh\left( \frac{H  dH/dt}{\nu}\right), \ \ \ \
\label{eq:l1_dhdt}
\end{eqnarray}
where $\lambda_{1 \uparrow}$, $\lambda_{1 \downarrow}$ and $\nu$ are parameters to be optimized.

This model is phenomenological, motivated by the empirical features of the data rather than by a microscopic theory of the underlying relaxation mechanisms.

\begin{figure}[tb]
\includegraphics[width=0.95\columnwidth]{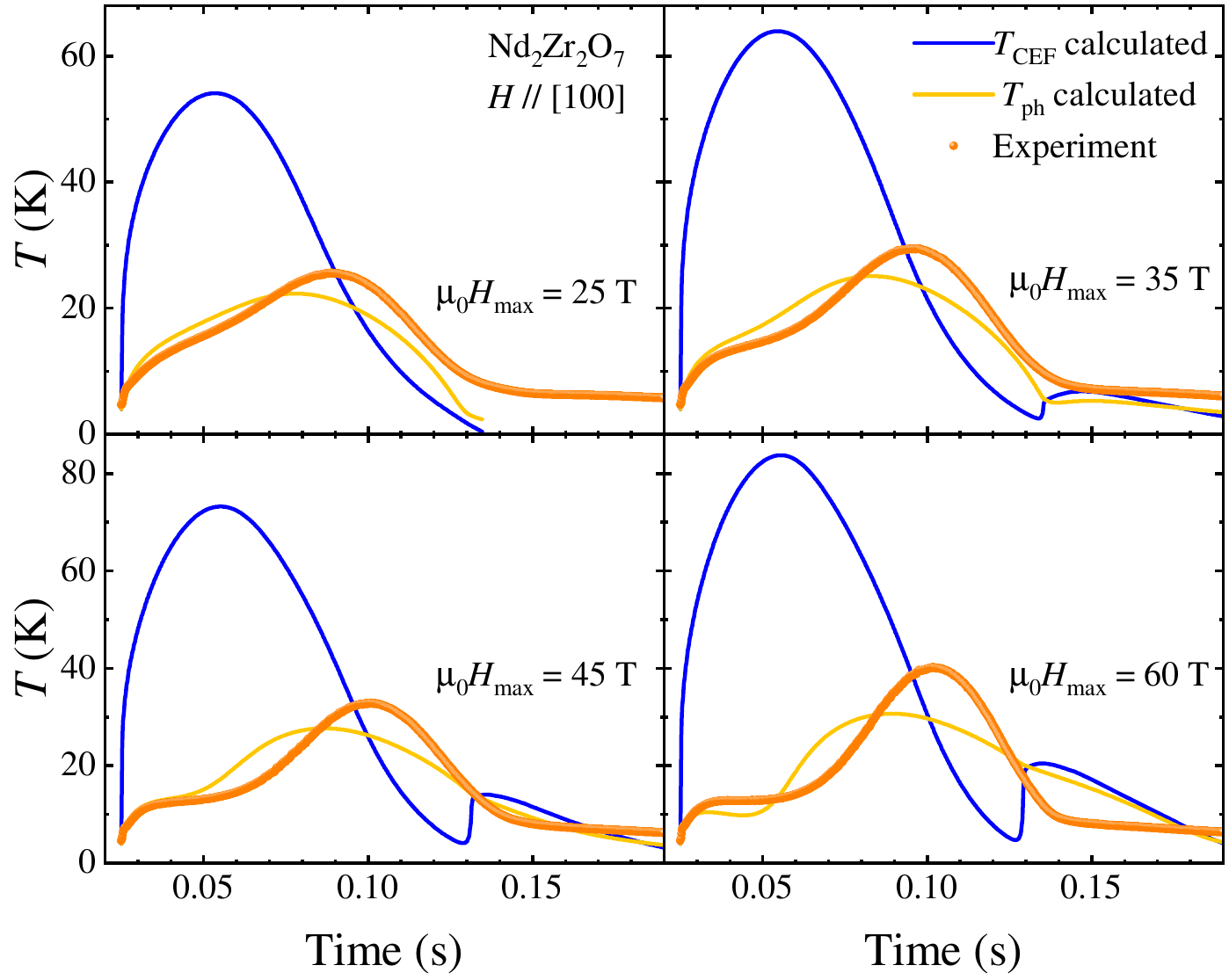}
\caption{Simulated crystal-field temperature $T_{\rm CEF} (t)$ and phonon temperature $T_{\rm ph} (t)$ with initial temperature $T_0=4$~K in \Nd, for a series of pulses with different maximum field strengths, compared with experimental data.
Calculations have been made using Eqs.~(\ref{eq:cef_time}) and (\ref{eq:phonontime}) and a pulse profile $H(t)$ reflecting that used in experiments (see Sec.~\ref{sec:experimental}), allowing the coupling $\lambda_1$ to depend on field sweep rate as in Eq. (\ref{eq:l1_dhdt}).
Calculations were made using model parameters given in Eq.~(\ref{eq:revised_model_params}).}
\label{fig:Nd_T(t)_refined_model}
\end{figure}

To optimize the parameters for the revised model, we used data from a series of field sweeps taken with initial temperature $T_0=4$~K.
Specifically, we simultaneously fit the data for pulses with maximum field strengths $\mu_0 H_{max} = 35,45$ and $60$~T.
Once parameters have been optimized we compare with the experimental results for all available field strengths - which includes also $\mu_0 H_{max} = 25$~T. 

The reason for not fitting to all field pulses simultaneously is that this results in worse fits to the higher field data, which contains the most interesting features that we wish to describe with the refined model.

The optimal parameters are:
\begin{eqnarray}
&&    \lambda_{1\uparrow} = 0.06~{\rm s}^{-1}, \ 
     \lambda_{1\downarrow} = 4.8~{\rm s}^{-1}, \nonumber \\
 &&     \lambda_{2} = 2.7~{\rm s}^{-1},  \ 
       \nu =  49000  \ T^{2}~{\rm s}^{-1}.
    \label{eq:revised_model_params}
\end{eqnarray}

The resulting comparisons with the experimental data are shown in Fig.~\ref{fig:Nd_T(t)_refined_model}.
While agreement with the data remains far from perfect, we now reproduce the plateau features visible for the stronger field pulses, while correctly retaining the absence of a plateau for weaker fields.
The size of the MCE is also more closely approximated with this model, although still underestimated.

\begin{figure}[tb]
\includegraphics[width=0.95\columnwidth]{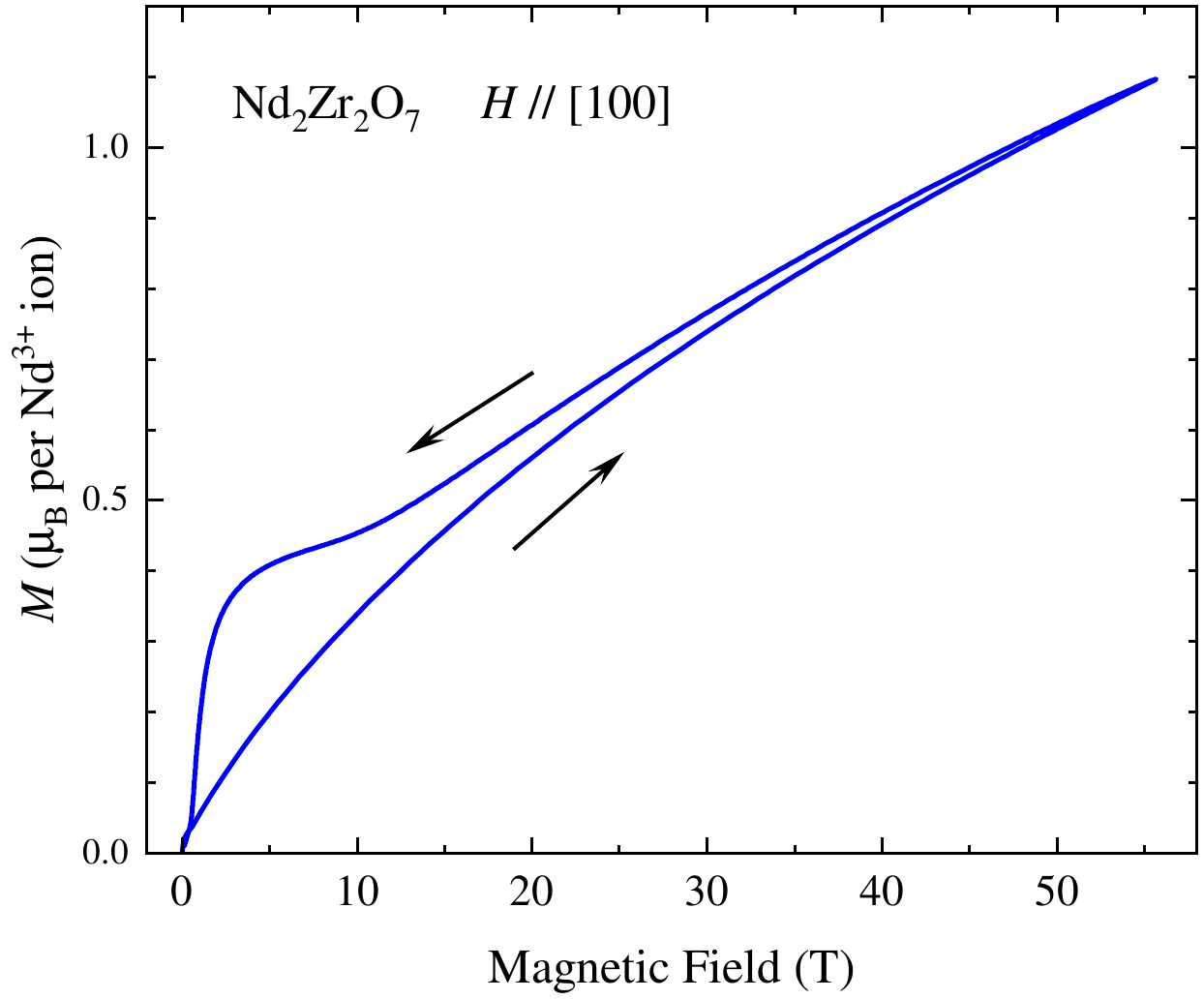}
\caption{Predicted magnetization curve for \Nd\ based on Eqs.~(\ref{eq:cef_time}) and~(\ref{eq:phonontime}), the sweep-rate dependence of the relaxation rate $\lambda_1$ given in Eq. (\ref{eq:l1_dhdt}), and the parameters given in Eq. (\ref{eq:revised_model_params}). 
Calculations are made assuming a starting temperature $T=4$~K and for field along the $[100]$ direction.
% The calculation correctly reproduces the lack of saturation even in large fields in the experimental data, as well as the larger magnetization on the down-sweep of the field, compared to the up-sweep, in spite of higher measured temperature.
% The kink observed at low fields in the calculation is not seen in the magnetization data at $T=4.2$~K but is observed at lower temperatures [see~Fig.~\ref{fig:Nd_MH_2_4K}].
% This calculation uses a faster field pulse $H(t)$ than the calculations of the MCE [Figs. \ref{fig:Nd_T(t)}-\ref{fig:Nd_T(H)}], to approximate the pulse used in the magnetization measurements (see Sec.~\ref{sec:experimental}).
}
\label{fig:Nd-magcurve}
\end{figure}

Using these same parameters, we have also calculated a predicted magnetization curve, based on the faster field pulse used in the magnetization experiments.
The results of these calculations are shown in Fig.~\ref{fig:Nd-magcurve}.
While agreement with the measured magnetization curve [Fig. \ref{fig:Nd_MH_2_4K}] is not perfect - in particular the theory predicts lower values of magnetization than are observed at high fields, a couple of key features of the data are captured.
The absence of any saturation in the magnetization, even at very large external fields, is reproduced by our analyses, and is due to the field-induced mixing of higher crystal-field levels with the ground doublet.
The sense of the observed hysteresis is also correct, with larger magnetizations predicted on the down sweep of the field, despite higher temperatures in the direct temperature measurements.
This is due to the loss of equilibrium between lattice and crystal field DOFs on the timescale of the experiment.

The kink observed at low fields in the calculation is not seen in the magnetization data at $T=4.2$~K but is observed at lower temperatures [see~Fig.~\ref{fig:Nd_MH_2_4K}].

%_________________________________________________________________________________________________
\section{Summary} \label{sec:summary}

In this work, we have reported experimental magnetization and MCE data, obtained using pulsed magnetic fields, for the pyrochlore zirconates \Nd\ and \Pra.
We find that both materials exhibit a strikingly large MCE.
This effect is strongly hysteretic, with significant differences between up and down sweeps of the applied field, even when the initial temperature lies well above any equilibrium magnetic ordering transition.

A notable feature of the data is the apparent inconsistency between  magnetization and temperature measurements when interpreted under equilibrium assumptions.
Specifically, the sample temperature is consistently higher on the down sweep of the field, which would suggest a reduced magnetization.
In contrast, we observe that the magnetization is larger on the down-sweep.
We interpret this counterintuitive behavior as evidence of a breakdown of thermal equilibrium between the magnetic and lattice subsystems on the timescale of the pulsed field experiment.

The time-domain data for \Nd\ reveal an additional striking signature: a long plateau-like feature in the measured temperature, $T(t)$, that terminates at the maximum field value.
This plateau is reproduced across multiple field pulses with different maximum amplitudes, suggesting that it is related to 
the change in sign of field-sweep-rate $dH/dt$, rather than being related to a particular value of $H$.

To rationalize these observations, we introduced a phenomenological two-subsystem model, in which the magnetic and lattice degrees of freedom are represented as weakly coupled subsystems, each characterized by its own effective temperature. With constant relaxation rates, this model accounts reasonably well for the data obtained in \Pra, capturing the gross features of both $T(H)$ and $M(H)$. However, it fails to describe the plateau in $T(t)$ that is characteristic of \Nd.

To model the \Nd\ results, we extended the framework by allowing the inter-subsystem thermal coupling to depend explicitly on the rate of change of the applied field. This refinement successfully reproduces the existence of the plateau, but the agreement with experiment in the detailed shapes of $T(H)$ and $M(H)$ curves remains limited.

Taken together, our findings demonstrate that the MCE in these pyrochlores is governed by  non-equilibrium dynamics, in which magnetic and lattice degrees of freedom exchange energy in a sweep-rate-dependent manner. 
The microscopic mechanism behind this behavior remains an open question. 

Future experiments should explore the implications of the apparent sweep-rate dependence of the thermal relaxation.
Experiments that could probe this include rotating the sample relative to an applied field~\cite{Kittaka_2018,Orendac_2018} or using low-frequency oscillating fields~\cite{Tokiwa_2011}.
If this unusual, non-equilibrium MCE can be understood, it may open up a new avenue to control heat flows in insulating materials.

%_________________________________________________________________________________________________
\section*{ACKNOWLEDGMENTS}
O.A.P. and O.B. acknowledge useful 
conversations with Claudio
Castelnovo.
The work at the University of Warwick was supported by EPSRC through grants EP/M028771/1 and EP/T005963/1.
O.A.P. acknowledges the EPSRC grant EP/X020304/1 that sponsored the secondment at the EMFL.
We also acknowledge support of the HLD at HZDR, member of the European Magnetic Field Laboratory (EMFL).
This work of Y.S., Y.G., S.Z., A.M, and J.W. has been supported in part by the DFG through SFB 1143 (project-id 247310070) and the W\"{u}rzburg-Dresden Cluster of Excellence on Complexity and Topology in Quantum Matter--$ct.qmat$ (EXC 2147, Project No.\ 390858490).

\appendix

\section{Calculation of thermodynamics from CEF spectrum}
\label{app:therm_cef}

In both \Nd\ and \Pra\ the CEF spectrum of the rare-earth ions features a ground-state doublet separated from higher energy levels by a large gap.
Therefore, for temperatures smaller than the gap ($\sim 100$~K), but larger than the scale of magnetic interactions ($\sim 1$~K), the entropy in the CEF degrees of freedom is close to $R \ln(2)$. 
Since, the rare-earth ions have substantial magnetic moments, a gap between the two states of the ground-state doublet opens rapidly with increasing field.
At fixed temperature, this would drive the entropy down from $\sim R \ln(2)$ towards zero.
In low fields, we therefore expect a large, negative value of $\left( \frac{\partial S}{\partial H} \right)_T$.

When this is combined with the smallness of the heat capacity of the CEF states, which is implied by the large gap to excited levels, this leads us to expect that $ \frac{d T_{\rm CEF}}{dt} $ will be large under adiabatic conditions [see Eq.~(\ref{eq:cef_time})], unless heat can be rapidly shared with the lattice degrees of freedom.

The CEF spectra for \Nd\ and \Pra\ can be modeled using the Hamiltonian:
\begin{eqnarray}
    \mathcal{H}_{CEF} = \sum_{nm} B_{nm} O_{nm},
    \label{eq:Hcef}
\end{eqnarray}
where $O_{nm}$ are Stevens operators~\cite{Stevens_1952, Hutchings_1964} and $B_{nm}$ are parameters determined by fitting to spectroscopic experiments, subject to restrictions imposed by the symmetry of the rare-earth site.

The parameters $B_{nm}$ have been determined for \Nd\ in \cite{Xu_2015} and for \Pra\ in \cite{Bonville_2016}, and we use these parameters in our model (see Appendix \ref{app:cef_details}).
To the CEF Hamiltonian we add a Zeeman coupling:
\begin{eqnarray}
    &&\mathcal{H} = \mathcal{H}_{CEF} + \mathcal{H}_{Z} 
    \label{eq:Htotal}
    \\
    && \mathcal{H}_{Z} = - g_J \mu_B {\bf H} \cdot {\bf J},
    \label{eq:HZ}
\end{eqnarray}
where $g_J$ is the Land{\'e} g-factor for the rare-earth ion ($8/11$ for Nd$^{3+}$, $4/5$ for Pr$^{3+}$), $\mu_B$ is the Bohr magneton and ${\bf J}$ is the total angular-momentum operator. The Python package {\it PyCrystalField} \cite{Scheie_2021} was used to implement the Hamiltonian based on the published values of $B_{nm}$.

With the definitions given in 
Eqs. (\ref{eq:Htotal})-(\ref{eq:HZ}) the Hamiltonian always has zero trace:
$$
\rm{Tr}[\mathcal{H}]=0.
$$
This condition defines the zero of energy
relative to which the eigenvalues are defined. 
Since this condition is true independently of applied field this choice of energy zero allows us to meaningfully differentiate eigenvalues with respect to field. 

We diagonalize this Hamiltonian for a series of values of
${\bf H}$ along the $[100]$ direction [Figs. \ref{fig:Pr_spec} and \ref{fig:Nd_spec}].
We then fit the dependence of the spectra on field strength to a second-order polynomial:
\begin{eqnarray}
    \omega_n (H) = \omega_{n0} + \alpha_{n1} H + \alpha_{n2} H^2.
    \label{eq:energies_approx}
\end{eqnarray}
The reason for performing these fits is that it gives us the functional dependence of the energy levels on magnetic field in an easily differentiable form, which is useful for computing $\left( \frac{\partial S}{\partial H}\right)_T$.
Because we incorporate the second-order term $\alpha_{n2}$, our model incorporates the mixing of crystal-field states at high magnetic fields.

For fields aligned along the $[100]$ direction, the spectrum thus produced is the same for all four sites in the unit cell.

For simplicity, from this point forward we keep only the four lowest-lying states for \Nd\ (corresponding to the ground-state and first-excited doublets in zero field) and the three lowest-lying states for \Pra\ (ground doublet and first-excited singlet).
The occupation of higher levels is negligible in any case at the temperatures and fields we consider.

The CEF heat capacity is then:
\begin{widetext}
    \begin{eqnarray}
    C_{\rm CEF} (H, T) = \frac{R}{k_{\rm B}^2 T^2}  \left( \frac{1}{Z(H, T) }  \sum_n \omega_n (H) ^2 \exp(-\omega_n (H) / k_{\rm B} T) - 
        \frac{1}{Z(H, T)^2} \left( \sum_n \omega_n (H)  \exp(-\omega_n (H) / k_{\rm B} T)\right)^2
    \right) \nonumber \\
\end{eqnarray}
\end{widetext}
where $Z(H, T)$ is the partition function:
\begin{eqnarray}
    Z (H, T) = \sum_n  \exp(-\omega_n (H) / k_{\rm B} T).
\end{eqnarray}

To calculate the entropy per mole we use the formula
\begin{eqnarray}
    S (H, T) = - R \sum_n p_n \ln(p_n),
\end{eqnarray}
where $p_n$ are the Boltzmann probabilities:
\begin{eqnarray}
   p_n = \frac{\exp(-\omega_n(H)/ k_{\rm B} T)}{Z(H, T)}.
\end{eqnarray}

Using this, we find that
\begin{eqnarray}
&&\left( \frac{\partial S}{\partial H} \right)_T =
-R \sum_n (1 + \ln(p_n)) \frac{\partial p_n}{\partial_H}\nonumber\\ 
&& = -R \sum_n \ln(p_n) \frac{\partial p_n}{\partial_H} \nonumber \\\
&&=R \sum_n
    \ln(p_n) \left(\frac{1}{Z k_{\rm B} T} \frac{\partial \omega_n}{\partial H} + \frac{1}{Z} \frac{\partial Z}{\partial H} \right)
   \nonumber \\
   &&=- \frac{1}{k_{\rm B} T} (S \langle M \rangle + \sum_n p_n \ln(p_n) \frac{\partial \omega_n}{\partial H}, \ \
\end{eqnarray}
where $\langle M \rangle$ is the average magnetization.
Using Eq. (\ref{eq:energies_approx}) gives
\begin{eqnarray}
&&\left( \frac{\partial S}{\partial H} \right)_T 
=- \frac{1}{k_{\rm B} T} (S \langle M \rangle + R \sum_n p_n \ln(p_n) (\alpha_{n1} + 2 \alpha_{n2} H)). \ \nonumber \\
\label{eq:dSdH-final}
\end{eqnarray}

All terms entering Eqs.~(\ref{eq:cef_time}) and (\ref{eq:phonontime}) are thus determined, with the exception of the relaxation rates $\lambda_1$ and $\lambda_2$ which we treat as adjustable parameters.

When calculating the magnetization curves, we use the following formula:
\begin{eqnarray}
    \langle M \rangle = 
    \sum_n p_n \frac{\partial \omega_n}{\partial H},
\end{eqnarray}
with the derivative calculated from the 
polynomial fit to $\omega_n(H)$.

\section{Details of crystal field model}
\label{app:cef_details}

Here we provide numerical details of the CEF model for \Nd\ and \Pra.

The Hamiltonian in zero field is given by Eq. (\ref{eq:Hcef}).
The symmetry of the point group at the rare-earth site causes a large number of terms to vanish.
The Stevens operators $O_{nm}$ associated to the nonvanishing terms are given below in terms of components of the angular momentum operators $J_{\alpha}$, with $\alpha=x,y,z$. Here, we are using local coordinates so that the $z$-axis is aligned with the local $C_3$ symmetry axis on each site.
\begin{eqnarray}
&& O_{20} = 3 J_z^2 - X  \nonumber \\
&& O_{40} = 35 J_z^4 - 30(X -25) J_z^2  + 3 X^2 - 6 X \nonumber \\
&& O_{43} = -\frac{1}{4} \{ J_+^3 + J_-^3, J_z \} \nonumber \\
&& O_{60} =  231 J_z^6 - (315X - 735) J_z^4 + \nonumber \\
&&  \qquad (105X^2 - 525X + 294) J_z^2 - 5X^3 + 40X^2 - 60X  \nonumber \\
&& O_{63} = -\frac{1}{4}  \left\{ (J_+^3 + J_-^3),(11J_z^3 - (3X + 59)J_z) \right\} \nonumber \\
&& O_{66} = \frac{1}{2} \left( J_+^6 + J_-^6 \right),
\end{eqnarray}
where
\begin{eqnarray}
X = J(J+1)
\end{eqnarray}
and $\{...\}$ is the anticommutator.

Xu {\it et al.}~\cite{Xu_2015} determined the following crystal-field parameters $B_{nm}$ for \Nd, which we adopt in this paper:
\begin{eqnarray}
&& B_{20} = -0.158~{\rm meV}, \ B_{40} = -0.015~{\rm meV}, \nonumber \\
&& B_{43} = -0.105~{\rm meV}, \ B_{60} = -0.000352~{\rm meV}, \nonumber \\
&& B_{63}  = 0.00477~{\rm meV}, \ B_{66} = -0.00502~{\rm meV}.
\end{eqnarray}

Implementing the Hamiltonian [Eq.~(\ref{eq:Hcef})] with these parameters and $J=9/2$, leads to a spectrum of 5 Kramers doublets, with the gaps between the 4 excited doublets and the ground state being:
\begin{eqnarray}
&&\Delta_1 = 23.34 \ {\rm meV}, \Delta_2 = 42.20 \ {\rm meV}, \nonumber \\
&&\Delta_3 = 43.59 \ {\rm meV}, \Delta_4 = 108.23 \ {\rm meV}. 
\end{eqnarray}
The size of these gaps differs slightly from the values reported in Ref.~[\onlinecite{Xu_2015}] because we neglect mixing between $J=9/2$ and $J=11/2$ manifolds for simplicity.
This mixing does not have a significant affect on the low-temperature behavior.

For \Pra, we follow the parameterization provided by Bonville {\it et al.}~\cite{Bonville_2016}:
\begin{eqnarray}
&& B_{20} = -0.631~{\rm meV}, \ 
B_{40} = -0.03236~{\rm meV}, \nonumber \\
&& B_{43} = -0.4674~{\rm meV}, \ 
B_{60} = -0.000245~{\rm meV}, \nonumber \\
&& B_{63}  = 0.001464~{\rm meV}, \ 
B_{66} = -0.001907~{\rm meV}.  
\end{eqnarray}

This results in a ground-state CEF doublet with a spectrum of excitations consisting of three singlets ($\Delta_{1,3,5}$) and two doublets ($\Delta_{2,4}$).
The calculated gaps when we neglect mixing with other $J$ manifolds are:
\begin{eqnarray}
&&\Delta_1 = 9.97 \ \rm{meV}, \Delta_2 = 56.90 \ \rm{meV}, \nonumber\\
&&\Delta_3 = 81.93\ \rm{meV}, \Delta_4 = 93.11 \ \rm{meV}, \nonumber \\
&&\Delta_5 = 107.69\ \rm{meV}. 
\end{eqnarray}
\bibliography{pyros_all}

@article{Skourski_2011,
  title = {High-field magnetization of {H}o$_2${F}e$_{17}$},
  author = {Skourski, Y. and Kuz'min, M. D. and Skokov, K. P. and Andreev, A. V. and Wosnitza, J.},
  journal = {Phys. Rev. B},
  volume = {83},
  issue = {21},
  pages = {214420},
  numpages = {9},
  year = {2011},
  month = {Jun},
  publisher = {American Physical Society},
  doi = {10.1103/PhysRevB.83.214420},
  url = {https://link.aps.org/doi/10.1103/PhysRevB.83.214420}}

@article{Kimura_2013,
	Author = {K. Kimura and S. Nakatsuji and J.-J. Wen and C. Broholm and M. B. Stone and E. Nishibori and H. Sawa},
	Journal = {Nat. Commun.},
	Pages = {1934},
	Title = {Quantum fluctuations in spin-ice-like {P}r$_2${Z}r$_2${O}$_7$},
	Volume = {4},
	Year = {2013}}

@article{Kimura_2013_JKPS,
Author = {Kimura, Kenta and Nakatsuji, Satoru and Nugroho, A. Agung},
Title = {Single-crystal study on the low-temperature magnetism of the pyrochlore magnet {P}r$_2${Z}r$_2${O}$_7$},
Journal = {J. Korean Phys. Soc.},
Year = {2013},
Volume = {63},
Number = {3, SI},
Pages = {719-721},
Month = {AUG},
DOI = {10.3938/jkps.63.719},
ISSN = {0374-4884},
EISSN = {1976-8524},
ResearcherID-Numbers = {Nugroho, Agustinus Agung/E-5977-2010},
ORCID-Numbers = {Nugroho, Agustinus Agung/0000-0002-1785-4008 Nakatsuji, Satoru/0000-0001-9134-659X},
Unique-ID = {WOS:000323502800108}}

@article{Kihara_2013,
    author = {Kihara, T. and Kohama, Y. and Hashimoto, Y. and Katsumoto, S. and Tokunaga, M.},
    title = {Adiabatic measurements of magneto-caloric effects in pulsed high magnetic fields up to 55 {T}},
    journal = {Rev. Sci. Instrum.},
    volume = {84},
    number = {7},
    pages = {074901},
    year = {2013},
    month = {07},
    issn = {0034-6748},
    doi = {10.1063/1.4811798}}

@article{Hatnean_2014,
	Author = {M. Ciomaga Hatnean and C. Decorse and M. R. Lees and O. A. Petrenko and D. S. Keeble and G. Balakrishnan},
	Journal = {Mater. Res. Express},
	Title = {Structural and Magnetic Properties of Single Crystals of the Geometrically Frustrated Zirconium Pyrochlore, {P}r$_2${Z}r$_2${O}$_7$},
    volume = {1},
    number = {2},
    pages = {026109},
    doi = {10.1088/2053-1591/1/2/026109},
	Year = {2014}}

@article{Erfanifam_2014,
  title = {Ultrasonic investigations of the spin ices {D}y$_2${T}i$_2${O}$_7$ and {H}o$_2${T}i$_2${O}$_7$ in and out of equilibrium},
  author = {Erfanifam, S. and Zherlitsyn, S. and Yasin, S. and Skourski, Y. and Wosnitza, J. and Zvyagin, A. A. and McClarty, P. and Moessner, R. and Balakrishnan, G. and Petrenko, O. A.},
  journal = {Phys. Rev. B},
  volume = {90},
  issue = {6},
  pages = {064409},
  numpages = {11},
  year = {2014},
  month = {Aug},
  publisher = {American Physical Society},
  doi = {10.1103/PhysRevB.90.064409},
  url = {https://link.aps.org/doi/10.1103/PhysRevB.90.064409}}

@article{Lhotel_2015,
  title = {Fluctuations and All-In--All-Out Ordering in Dipole-Octupole {N}d$_2${Z}r$_2${O}$_7$},
  author = {Lhotel, E. and Petit, S. and Guitteny, S. and Florea, O. and Ciomaga Hatnean, M. and Colin, C. and Ressouche, E. and Lees, M. R. and Balakrishnan, G.},
  journal = {Phys. Rev. Lett.},
  volume = {115},
  issue = {19},
  pages = {197202},
  numpages = {5},
  year = {2015},
  month = {Nov},
  publisher = {American Physical Society},
  doi = {10.1103/PhysRevLett.115.197202},
  url = {https://link.aps.org/doi/10.1103/PhysRevLett.115.197202}}

@article{Ciomaga_2016,
	AUTHOR = {Ciomaga Hatnean, Monica and Decorse, Claudia and Lees, Martin R. and Petrenko, Oleg A. and Balakrishnan, Geetha},
	TITLE = {Zirconate Pyrochlore Frustrated Magnets: Crystal Growth by the Floating Zone Technique},
	JOURNAL = {Crystals},
	VOLUME = {6},
	YEAR = {2016},
	NUMBER = {7},
	pages = {79},
	URL = {https://www.mdpi.com/2073-4352/6/7/79},
	ISSN = {2073-4352},
	DOI = {10.3390/cryst6070079}}

@article{Petit_2016,
	Author = {Petit, S. and Lhotel, E. and Canals, B. and Hatnean, M. Ciomaga and Ollivier, J. and Mutka, H. and Ressouche, E. and Wildes, A. R. and Lees,  M. R. and Balakrishnan, G.},
	Title = {Observation of magnetic fragmentation in spin ice},
	Journal = {Nat. Phys.},
	Year = {2016},
	Volume = {12},
	Number = {8},
	Pages = {746-750},
	Month = {AUG},
	DOI = {10.1038/NPHYS3710},
	ISSN = {1745-2473},
	URL  =  {https://doi.org/10.1038/nphys3710},
	EISSN = {1745-2481}}

@article{Rau_2019,
  author       = {Rau, Jeffrey G. and Gingras, Michel J. P.},
  title        = {Frustrated Quantum Rare-Earth Pyrochlores},
  journal      = {Annu. Rev. Condens. Matter Phys.},
  volume       = {10},
  number       = {},
  pages        = {357--386},
  year         = {2019},
  doi          = {10.1146/annurev-conmatphys-022317-110520},
  url          = {https://doi.org/10.1146/annurev-conmatphys-022317-110520}}

@article{Benton_2016,
  title = {Quantum origins of moment fragmentation in $\mathrm{Nd_2Zr_2O_7}$},
  author = {Benton, Owen},
  journal = {Phys. Rev. B},
  volume = {94},
  issue = {10},
  pages = {104430},
  numpages = {6},
  year = {2016},
  month = {Sep},
  publisher = {American Physical Society},
  doi = {10.1103/PhysRevB.94.104430},
  url = {https://link.aps.org/doi/10.1103/PhysRevB.94.104430}}

@article{Bonville_2016,
  title = {Magnetic properties and crystal field in {P}r$_2${Z}r$_2${O}$_7$},
  author = {Bonville, P. and Guitteny, S. and Gukasov, A. and Mirebeau, I. and Petit, S. and Decorse, C. and Hatnean, M. Ciomaga and Balakrishnan, G.},
  journal = {Phys. Rev. B},
  volume = {94},
  issue = {13},
  pages = {134428},
  numpages = {7},
  year = {2016},
  month = {Oct},
  publisher = {American Physical Society},
  doi = {10.1103/PhysRevB.94.134428},
  url = {https://link.aps.org/doi/10.1103/PhysRevB.94.134428}}

@article{Wen_2017,
  title = {Disordered Route to the {C}oulomb Quantum Spin Liquid: Random Transverse Fields on Spin Ice in $\mathrm{Pr_2Zr_2O_7}$},
  author = {Wen, J.-J. and Koohpayeh, S. M. and Ross, K. A. and Trump, B. A. and McQueen, T. M. and Kimura, K. and Nakatsuji, S. and Qiu, Y. and Pajerowski, D. M. and Copley, J. R. D. and Broholm, C. L.},
  journal = {Phys. Rev. Lett.},
  volume = {118},
  issue = {10},
  pages = {107206},
  numpages = {5},
  year = {2017},
  month = {Mar},
  publisher = {American Physical Society},
  doi = {10.1103/PhysRevLett.118.107206},
  url = {https://link.aps.org/doi/10.1103/PhysRevLett.118.107206}}

@article{Brambleby_2017,
  title = {Adiabatic physics of an exchange-coupled spin-dimer system: Magnetocaloric effect, zero-point fluctuations, and possible two-dimensional universal behavior},
  author = {Brambleby, J. and Goddard, P. A. and Singleton, J. and Jaime, M. and Lancaster, T. and Huang, L. and Wosnitza, J. and Topping, C. V. and Carreiro, K. E. and Tran, H. E. and Manson, Z. E. and Manson, J. L.},
  journal = {Phys. Rev. B},
  volume = {95},
  issue = {2},
  pages = {024404},
  numpages = {12},
  year = {2017},
  month = {Jan},
  publisher = {American Physical Society},
  doi = {10.1103/PhysRevB.95.024404},
  url = {https://link.aps.org/doi/10.1103/PhysRevB.95.024404}}

@article{Lhotel_2018,
	Author = {Lhotel, E. and Petit, S. and Hatnean, M. Ciomaga and Ollivier, J. and Mutka, H. and Ressouche, E. and Lees, M. R. and Balakrishnan, G.},
	Title = {Evidence for dynamic kagome ice},
	Journal = {Nat. Commun.},
	Year = {2018},
	Volume = {9},
	Month = {SEP 17},
	DOI = {10.1038/s41467-018-06212-2},
	pages = {3786},
	URL  =  {https://doi.org/10.1038/s41467-018-06212-2},
	SSN = {2041-1723}}

@article{Opherden_2019,
  title = {Magnetization beyond the {I}sing limit of {H}o$_2${T}i$_2${O}$_7$},
  author = {Opherden, L. and Herrmannsd\"orfer, T. and Uhlarz, M. and Gorbunov, D. I. and Miyata, A. and Portugall, O. and Ishii, I. and Suzuki, T. and Kaneko, H. and Suzuki, H. and Wosnitza, J.},
  journal = {Phys. Rev. B},
  volume = {99},
  issue = {8},
  pages = {085132},
  numpages = {6},
  year = {2019},
  month = {Feb},
  publisher = {American Physical Society},
  doi = {10.1103/PhysRevB.99.085132},
  url = {https://link.aps.org/doi/10.1103/PhysRevB.99.085132}}

@article{Smith_2022,
  title = {Case for a $\mathrm{U(1)_\pi}$ Quantum Spin Liquid Ground State in the Dipole-Octupole Pyrochlore $\mathrm{Ce_2Zr_2O_7}$},
  author = {Smith, E. M. and Benton, O. and Yahne, D. R. and Placke, B. and Sch\"afer, R. and Gaudet, J. and Dudemaine, J. and Fitterman, A. and Beare, J. and Wildes, A. R. and Bhattacharya, S. and DeLazzer, T. and Buhariwalla, C. R. C. and Butch, N. P. and Movshovich, R. and Garrett, J. D. and Marjerrison, C. A. and Clancy, J. P. and Kermarrec, E. and Luke, G. M. and Bianchi, A. D. and Ross, K. A. and Gaulin, B. D.},
  journal = {Phys. Rev. X},
  volume = {12},
  issue = {2},
  pages = {021015},
  numpages = {19},
  year = {2022},
  month = {Apr},
  publisher = {American Physical Society},
  doi = {10.1103/PhysRevX.12.021015},
  url = {https://link.aps.org/doi/10.1103/PhysRevX.12.021015}}

@article{Tang_2023,
   Author = {Tang, Nan and Gritsenko, Yulia and Kimura, Kenta and Bhattacharjee, Subhro and Sakai, Akito and Fu, Mingxuan and Takeda, Hikaru and Man, Huiyuan and Sugawara, Kento and Matsumoto, Yosuke and Shimura, Yasuyuki
   and Wen, Jiajia and Broholm, Collin and Sawa, Hiroshi and Takigawa, Masashi and Sakakibara, Toshiro and Zherlitsyn, Sergei and Wosnitza, Joachim and Moessner, Roderich and Nakatsuji, Satoru},
   Title = {Spin-orbital liquid state and liquid-gas metamagnetic transition on a pyrochlore lattice},
  Journal = {Nat. Phys.},
 Year = {2023},
 Volume = {19},
 Number = {1},
 Pages = {92 - 98},
 Month = {JAN},
DOI = {10.1038/s41567-022-01816-4},
EarlyAccessDate = {DEC 2022},
ISSN = {1745-2473},
EISSN = {1745-2481}}

@article{Tang_2024,
  title = {Crystal field magnetostriction of spin ice under ultrahigh magnetic fields},
  author = {Tang, Nan and Gen, Masaki and Rotter, Martin and Man, Huiyuan and Matsuhira, Kazuyuki and Matsuo, Akira and Kindo, Koichi and Ikeda, Akihiko and Matsuda, Yasuhiro H. and Gegenwart, Philipp and Nakatsuji, Satoru and Kohama, Yoshimitsu},
  journal = {Phys. Rev. B},
  volume = {110},
  issue = {21},
  pages = {214414},
  numpages = {10},
  year = {2024},
  month = {Dec},
  publisher = {American Physical Society},
  doi = {10.1103/PhysRevB.110.214414},
  url = {https://link.aps.org/doi/10.1103/PhysRevB.110.214414}}

@article{Benton_2018,
  title = {Instabilities of a {U}(1) Quantum Spin Liquid in Disordered Non-{K}ramers Pyrochlores},
  author = {Benton, Owen},
  journal = {Phys. Rev. Lett.},
  volume = {121},
  issue = {3},
  pages = {037203},
  numpages = {7},
  year = {2018},
  month = {Jul},
  publisher = {American Physical Society},
  doi = {10.1103/PhysRevLett.121.037203},
  url = {https://link.aps.org/doi/10.1103/PhysRevLett.121.037203}}

@article{Stevens_1952,
  author    = {K. W. H. Stevens},
  title     = {Matrix Elements and Operator Equivalents Connected with the Magnetic Properties of Rare Earth Ions},
  journal   = {Proc. Phys. Soc. London A},
  year      = {1952},
  volume    = {65},
  number    = {3},
  pages     = {209--215},
  doi       = {10.1088/0370-1298/65/3/308}}

@incollection{Hutchings_1964,
    title = {Point-Charge Calculations of Energy Levels of Magnetic Ions in Crystalline Electric Fields},
    editor = {Frederick Seitz and David Turnbull},
    booktitle = {Solid State Physics},
    publisher = {Academic Press},
    volume = {16},
    pages = {227-273},
    year = {1964},
    issn = {0081-1947},
    doi = {https://doi.org/10.1016/S0081-1947(08)60517-2},
    url = {https://www.sciencedirect.com/science/article/pii/S0081194708605172},
    author = {M.T. Hutchings}}

@article{Xu_2015,
  title = {Magnetic structure and crystal-field states of the pyrochlore antiferromagnet $\mathrm{Nd_2Zr_2O_7}$},
  author = {Xu, J. and Anand, V. K. and Bera, A. K. and Frontzek, M. and Abernathy, D. L. and Casati, N. and Siemensmeyer, K. and Lake, B.},
  journal = {Phys. Rev. B},
  volume = {92},
  issue = {22},
  pages = {224430},
  numpages = {12},
  year = {2015},
  month = {Dec},
  publisher = {American Physical Society},
  doi = {10.1103/PhysRevB.92.224430},
  url = {https://link.aps.org/doi/10.1103/PhysRevB.92.224430}}

@article{Kittaka_2018,
author = {Kittaka, Shunichiro and Nakamura, Shota and Kadowaki, Hiroaki and Takatsu, Hiroshi and Sakakibara,Toshiro},
title = {Field-rotational Magnetocaloric Effect: A New Experimental Technique for Accurate Measurement of the Anisotropic Magnetic Entropy},
journal = {J. Phys. Soc. Jpn.},
volume = {87},
number = {7},
pages = {073601},
year = {2018},
doi = {10.7566/JPSJ.87.073601}}

@article{Orendac_2018,
author = {Orendac, Matus and Gabani, Slavomir and Gazo, Emil and Pristas, Gabriel and Shitsevalova, Natalya and Siemensmeyer, Konrad and Flachbart, Karol},
title = {Rotating magnetocaloric effect and unusual magnetic features in metallic strongly anisotropic geometrically frustrated {T}m{B}$_4$},
journal = {Sci. Rep.},
volume = {8},
pages = {10933},
year = {2018},
doi = {/10.1038/s41598-018-29399-2}}

@article{Tokiwa_2011,
    author = {Tokiwa, Y. and Gegenwart, P.},
    title = {High-resolution alternating-field technique to determine the magnetocaloric effect of metals down to very low temperatures},
    journal = {Rev. Sci. Instrum.},
    volume = {82},
    number = {1},
    pages = {013905},
    year = {2011},
    month = {01},
    issn = {0034-6748},
    doi = {10.1063/1.3529433}}

@article{Gao_2019,
    author = {Gao, Bin and Chen, Tong and Tam, David and Huang, Chien-Lung and Sasmal, Kalyan and Adroja, Devashibhai and Ye, Feng and Cao, Huibo and Sala, Gabriele and Stone, Matthew and Baines, Christopher and Barker, Joel and Hu, Haoyu and Chung, Jae-Ho and Xu, Xianghan and Cheong, Sang-Wook and Nallaiyan, Manivannan and Spagna, Stefano and Maple, M. and Dai, Pengcheng},
    year = {2019},
    month = {07},
    pages = {1052–1057},
    title = {Experimental Signatures of a Three-dimensional Quantum Spin Liquid in Effective Spin-1/2 {C}e$_2${Z}r$_2${O}$_7$ Pyrochlore},
    volume = {15},
    journal = {Nat. Phys.},
    doi = {10.1038/s41567-019-0577-6}}

@article{Gaudet_2019,
    title = {Quantum Spin Ice Dynamics in the Dipole-Octupole Pyrochlore Magnet {C}e$_2${Z}r$_2${O}$_7$},
  author = {Gaudet, J. and Smith, E. M. and Dudemaine, J. and Beare, J. and Buhariwalla, C. R. C. and Butch, N. P. and Stone, M. B. and Kolesnikov, A. I. and Xu, Guangyong and Yahne, D. R. and Ross, K. A. and Marjerrison, C. A. and Garrett, J. D. and Luke, G. M. and Bianchi, A. D. and Gaulin, B. D.},
  journal = {Phys. Rev. Lett.},
  volume = {122},
  issue = {18},
  pages = {187201},
  numpages = {6},
  year = {2019},
  month = {May},
  publisher = {American Physical Society},
  doi = {10.1103/PhysRevLett.122.187201},
  url = {https://link.aps.org/doi/10.1103/PhysRevLett.122.187201}}

@article{Ciomaga_2015,
  title = {Structural and magnetic investigations of single-crystalline neodymium zirconate pyrochlore {N}d$_2${Z}r$_2${O}$_7$},
  author = {Hatnean, M. Ciomaga and Lees, M. R. and Petrenko, O. A. and Keeble, D. S. and Balakrishnan, G. and Gutmann, M. J. and Klekovkina, V. V. and Malkin, B. Z.},
  journal = {Phys. Rev. B},
  volume = {91},
  issue = {17},
  pages = {174416},
  numpages = {9},
  year = {2015},
  month = {May},
  publisher = {American Physical Society},
  doi = {10.1103/PhysRevB.91.174416},
  url = {https://link.aps.org/doi/10.1103/PhysRevB.91.174416}
}

@article{Singh_2008,
  title = {Manifestation of geometric frustration on magnetic and thermodynamic properties of the pyrochlores {S}m$_2${X}$_2${O}$_7$ ($\mathrm{X}=\mathrm{Ti},\mathrm{Zr}$)},
  author = {Singh, Surjeet and Saha, Surajit and Dhar, S. K. and Suryanarayanan, R. and Sood, A. K. and Revcolevschi, A.},
  journal = {Phys. Rev. B},
  volume = {77},
  issue = {5},
  pages = {054408},
  numpages = {7},
  year = {2008},
  month = {Feb},
  publisher = {American Physical Society},
  doi = {10.1103/PhysRevB.77.054408},
  url = {https://link.aps.org/doi/10.1103/PhysRevB.77.054408}
}

@phdthesis{Xu_2017_Thesis,
  author       = {Xu, J.},
  title        = {Magnetic Properties of Rare Earth Zirconate Pyrochlores},
  school       = {Technische Universit{\"a}t Berlin},
  year         = {2017},
  address      = {Berlin, Germany},
}

@article{Smith_2024,
  title = {Single-Crystal Diffuse Neutron Scattering Study of the Dipole-Octupole Quantum Spin-Ice Candidate {C}e$_2${Z}r$_2${O}$_7$: No Apparent Octupolar Correlations Above {T}=0.05~{K}},
  author = {Smith, E. M. and Sch\"afer, R. and Dudemaine, J. and Placke, B. and Yuan, B. and Morgan, Z. and Ye, F. and Moessner, R. and Benton, O. and Bianchi, A. D. and Gaulin, B. D.},
  journal = {Phys. Rev. X},
  volume = {15},
  issue = {2},
  pages = {021033},
  numpages = {20},
  year = {2025},
  month = {Apr},
  publisher = {American Physical Society},
  doi = {10.1103/PhysRevX.15.021033},
  url = {https://link.aps.org/doi/10.1103/PhysRevX.15.021033}
}

@article{Gao_2022,
  title = {Magnetic field effects in an octupolar quantum spin liquid candidate},
  author = {Gao, Bin and Chen, Tong and Yan, Han and Duan, Chunruo and Huang, Chien-Lung and Yao, Xu Ping and Ye, Feng and Balz, Christian and Stewart, J. Ross and Nakajima, Kenji and Ohira-Kawamura, Seiko and Xu, Guangyong and Xu, Xianghan and Cheong, Sang-Wook and Morosan, Emilia and Nevidomskyy, Andriy H. and Chen, Gang and Dai, Pengcheng},
  journal = {Phys. Rev. B},
  volume = {106},
  issue = {9},
  pages = {094425},
  numpages = {10},
  year = {2022},
  month = {Sep},
  publisher = {American Physical Society},
  doi = {10.1103/PhysRevB.106.094425},
  url = {https://link.aps.org/doi/10.1103/PhysRevB.106.094425}
}

@article{Gao_arxiv,
  title = {Emergent photons and fractionalized excitations in a quantum spin liquid},
  author = {Gao, Bin and Desrochers, F. and
            Tam, D. W. and Steffens, P. and
            Hiess, A. and Su, Y. and Cheong, S. W. and Kim, Y. B. and Dai, P.},
  journal = {arXiv:2404.04207},
  url = {https://arxiv.org/abs/2404.04207}
}

@article{Xu_2019,
  title = {Anisotropic exchange Hamiltonian, magnetic phase diagram, and domain inversion of {N}d$_2${Z}$_2${O}$_7$},
  author = {Xu, J. and Benton, Owen and Anand, V. K. and Islam, A. T. M. N. and Guidi, T. and Ehlers, G. and Feng, E. and Su, Y. and Sakai, A. and Gegenwart, P. and Lake, B.},
  journal = {Phys. Rev. B},
  volume = {99},
  issue = {14},
  pages = {144420},
  numpages = {14},
  year = {2019},
  month = {Apr},
  publisher = {American Physical Society},
  doi = {10.1103/PhysRevB.99.144420},
  url = {https://link.aps.org/doi/10.1103/PhysRevB.99.144420}
}

@article{Li_arxiv,
  title        = {Realization of discretized response in rare-earth vanadates accessed by AC susceptibility and magnetocaloric methods},
  author = {Yuntian Li and Linda Ye and Mark P. Zic and Arkady Shekhter and Ian R. Fisher},
  journal       = {arXiv:2502.16327},
  year         = {2025},
  url          = {https://arxiv.org/abs/2502.16327v3}}

@article{Sanders_1977,
  title = {Effect of magnon-phonon thermal relaxation on heat transport by magnons},
  author = {Sanders, D. J. and Walton, D.},
  journal = {Phys. Rev. B},
  volume = {15},
  issue = {3},
  pages = {1489--1494},
  numpages = {0},
  year = {1977},
  month = {Feb},
  publisher = {American Physical Society},
  doi = {10.1103/PhysRevB.15.1489},
  url = {https://link.aps.org/doi/10.1103/PhysRevB.15.1489}
}

@article{Liao_2014,
  title = {Generalized Two-Temperature Model for Coupled Phonon-Magnon Diffusion},
  author = {Liao, Bolin and Zhou, Jiawei and Chen, Gang},
  journal = {Phys. Rev. Lett.},
  volume = {113},
  issue = {2},
  pages = {025902},
  numpages = {6},
  year = {2014},
  month = {Jul},
  publisher = {American Physical Society},
  doi = {10.1103/PhysRevLett.113.025902},
  url = {https://link.aps.org/doi/10.1103/PhysRevLett.113.025902}
}

@article{Scheie_2021,
  author       = {Scheie, A.},
  title        = {{P}y{C}rystal{F}ield: software for calculation, analysis and fitting of crystal electric field {H}amiltonians},
  journal      = {J. Appl. Crystallogr.},
  volume       = {54},
  number       = {Pt 1},
  pages        = {356--362},
  year         = {2021},
  doi          = {10.1107/S160057672001554X},
  url          = {https://journals.iucr.org/paper?S160057672001554X}}

@Misc{scipy-ref,
  author =    {Eric Jones and Travis Oliphant and Pearu Peterson and others},
  title =     {{SciPy}: Open source scientific tools for {Python}},
  year =      {2001--},
  url = "http://www.scipy.org/"}
\end{document}